\documentclass[aip,reprint,twocolumn]{revtex4-2}

\usepackage{graphicx}
\usepackage{dcolumn}
\usepackage{bm}
\usepackage{color}
\usepackage{amsmath}
\usepackage{amssymb}
\usepackage{graphicx}
\usepackage[unicode=true,pdfusetitle,
 bookmarks=true,bookmarksnumbered=false,bookmarksopen=false,
 breaklinks=false,pdfborder={0 0 0},pdfborderstyle={},backref=false,colorlinks=true]
 {hyperref}
\hypersetup{
 pdfborderstyle={},citecolor=blue}

\usepackage[utf8]{inputenc}
\usepackage[T1]{fontenc}
\usepackage{mathptmx}

\begin{document}

 \title{Temporal localized states and square-waves in semiconductor micro-resonators with strong time-delayed feedback}
 \author{Elias R. Koch}
 \affiliation{Institute for Theoretical Physics, University of M\"unster, Wilhelm-Klemm-Str. 9, 48149 M\"unster, Germany}
 \author{Thomas G. Seidel}
 \affiliation{Institute for Theoretical Physics, University of M\"unster, Wilhelm-Klemm-Str. 9, 48149 M\"unster, Germany}
\author{Julien Javaloyes}
\affiliation{Departament de Física \& IAC-3, Universitat de les Illes Balears, C/ Valldemossa km 7.5, 07122 Mallorca, Spain}
\author{Svetlana V. Gurevich}
\email{gurevics@uni-muenster.de}
\affiliation{Institute for Theoretical Physics, University of M\"unster, Wilhelm-Klemm-Str. 9, 48149 M\"unster, Germany}
 \affiliation{Center for Nonlinear Science (CeNoS), University of M\"unster, Corrensstrasse 2, 48149 M\"unster, Germany}
 \affiliation{Departament de Física \& IAC-3, Universitat de les Illes Balears,
 C/ Valldemossa km 7.5, 07122 Mallorca, Spain}
\begin{abstract}
In this paper we study the dynamics of a vertically-emitting micro-cavity operated in the Gires-Tournois regime that contains a semiconductor quantum-well and that is subjected to strong time-delayed optical feedback and detuned optical injection. Using a first principle time-delay model for the optical response, we disclose sets of multistable dark and bright temporal localized states coexisting on their respective bistable homogeneous backgrounds. In the case of anti-resonant optical feedback, we identify square-waves with a period of twice the round-trip in the external cavity. Finally, we perform a multiple time-scale analysis in the good cavity limit. The resulting normal form is in good agreement with the original time-delayed model.
\end{abstract}

\maketitle
\textbf{ We theoretically study the dynamics of temporal localized states and square-waves in an injected micro-resonator enclosed into a long external cavity. Departing from former studies where only an intensity  dependent refractive index (i.e. a Kerr nonlinearity) was used, we consider a semiconductor quantum well as the nonlinear element. A first principle time-delayed model for the optical response is employed. By conducting extended direct numerical simulations and utilizing path-continuation techniques, we demonstrate the  existence of a bistable set of bright and dark temporally localized states as well as square-waves. To clarify the influence of the second and third order chromatic dispersion and of the relative position with respect to the band-gap frequency of the quantum well, we derive a normal form partial differential equation in the good cavity limit and compare its results with those obtained with the original time-delayed model.}

\section{Introduction}

Optical frequency combs (OFCs) denote light fields consisting of a discrete series of evenly spaced spectral components (or colors) that maintain high coherence across the whole bandwidth with ultra-low-noise~\cite{CY-RMP-03}. Their applications encompass a variety of fields including ultra-broadband coherent optical communications, high-precision optical spectroscopy, high precision metrology and optical arbitrary pulse generation~\cite{UHH-NAT-02,CW-NAP-10,D-JOSAB-10,PPR-PR-18}, to name just a few. Prominent examples of OFCs realizations leverage the temporal cavity solitons obtained in injected passive high-Q Kerr resonators, such as microrings \cite{HBJ-NAP-14} and fiber loops \cite{LCK-NAP-10} or employ active optical sources such as mode-locked lasers~\cite{tropper04,LMB-OE-10,WVM-LSP-17}.

Recently, the promising potential of injected Kerr micro-cavities enclosed into a long external feedback cavity as sources of high-power tunable OFCs, temporal localized structures (TLSs) and square-waves (SWs) was demonstrated~\cite{SPV-OL-19,SJG-OL-22,KSG-OL-22}. There, it was shown that the natural modeling approach consists in using singular time-delayed systems, in which the algebraic constraints simply corresponds to the boundary conditions linking the optical field in the various parts of the compound cavity system. In the normal dispersion regime and for the case of resonant optical feedback, dark and bright TLSs (and corresponding OFCs) appear via the locking of domain walls that connect the high and low bistable continuous wave (CW) intensity levels. Due to the oscillatory tails induced by the cavity dispersion~\cite{SCM-PRL-19}, these solutions can interlock at multiple equilibrium distances leading to a rich solution multistability. Moreover, operated out of the bistability regime and for anti-resonant feedback, SWs possessing a period of twice the round-trip in the external cavity were disclosed. Interestingly, the latter appear through a homoclinic snaking scenario leading to the formation of complex-shaped multistable solutions.

Departing from former studies, in this paper we analyze the impact of considering a semiconductor quantum well as the nonlinear element. A first principle model for the optical response is employed which allows to explore e.g., the influence of the detuning with respect to the semiconductor band-gap that controls the ratio between nonlinear index change and absorption. Using a combination of analytical, numerical and path-continuation methods, we shall show that this extended model predicts the existence of a multistable set of bright and dark TLSs coexisting on their respective bistable homogeneous CW backgrounds. Furthermore, in the case of anti-resonant optical feedback, the dynamics of resulting SWs is analyzed. Finally, in order to clarify the influence of the second and third order chromatic dispersion and of the frequency dependence of the quantum-well response, we perform a multiple time-scale analysis in the good cavity limit. The resulting dispersive normal form partial differential equation shows a good qualitative agreement with the original time-delayed model.

\section{Model equations}

\begin{figure}
	\includegraphics[width=1\columnwidth]{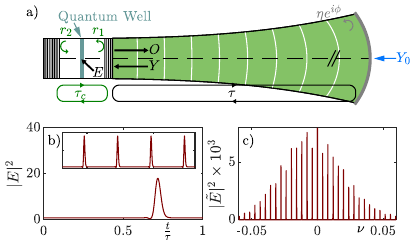}
	\caption{a) Schematic of a micro-cavity containing a quantum well layer coupled to an external cavity with round trip $\tau$ which is closed by a
mirror with reflectivity $\eta$ and phase $\phi$, and driven by a CW
beam with amplitude $Y_0$. b) A single TLS circulating in the external cavity obtained by a numerical integration of Eqs.~\eqref{eq:QGTI1}-\eqref{eq:QGTI3}. The inset shows the dynamics over several round-trips. c) Power density spectrum of the solution in b). Parameters are $(\delta,h,\eta,\varphi,\gamma,f,u,b,\tau)=(0.015,2,0.99,0,1,0.2,0.65,1000,250)$.} \label{fig:fig1}
	\end{figure}

The schematic setup is depicted in Fig.~\ref{fig:fig1}~(a). It is composed of a monomode micro-cavity of a few micrometers in length, with a radius up to $100\,\mathrm{\mu m}$ and with round-trip time $\tau_c$, which contains a thin quantum well layer acting as a nonlinear medium that is situated at the anti-node of the field. The micro-cavity is closed by two distributed Bragg mirrors with reflectivities $r_{1,2}$ and it is coupled
to a long external cavity of (typically) a few centimeters with
round-trip time $\tau\gg\tau_c$. The system is subjected to CW injection with
amplitude $Y_0$ and frequency $\omega_0$. The detuning with respect to the
micro-cavity resonance $\omega_c$ is $\delta=\omega_c-\omega_0$. The external cavity is closed by a feedback mirror with reflectivity $\eta$. The total external cavity phase $\varphi$ consists of the accumulated phase per round-trip due to propagation and the phase of the feedback mirror $\phi$. 

Following the methods developed in Refs.~\cite{MB-JQE-05,SPV-OL-19}, the first-principles model for the system shown in Fig.~\ref{fig:fig1}~(a) reads
\begin{align}
        \dot{E}&=\left[if\chi -1 -i\delta\right]E+hY\,, \label{eq:QGTI1}\\
        \dot{D}&=\gamma\left[-D+\Im(\chi)|E|^2\right]\,, \label{eq:QGTI2}\\
        Y&=\eta e^{i\varphi}\left[E(t-\tau)-Y(t-\tau)\right]+Y_0\sqrt{1-\eta^2}\,.\label{eq:QGTI3}
\end{align}
Here, $E$ and $Y$ denote the slowly varying envelope of electric
fields in the micro-cavity and the external cavity, respectively. Further, $D$ is the carrier density normalized to the transparency level at the band gap, $\gamma$ is the scaled carrier lifetime, and $f$ is an effective dimensionless gain prefactor. The latter is the product of a geometrical factor that contains the cavity enhancement effect and the active medium length, normalized by the photon lifetime in the micro-cavity. 
The field cavity enhancement can be conveniently scaled out using the Stokes relations allowing the fields $E$ and $Y$ to be of the same order of magnitude. This leads to a simple input-output relation $O = E - Y$: The output $O$ is the combination of the intra-cavity photons transmitted and reflected by the micro-cavity and it is re-injected after attenuation and a time delay $\tau$. The coupling between the intra- and external cavity fields is given by Eq.~\eqref{eq:QGTI3} which is a delay algebraic equation (DAE). The DAE~\eqref{eq:QGTI3} takes into account all the multiple reflections in a possibly high finesse external cavity where $\eta\lesssim 1$.

The light coupling efficiency in the cavity is given by the factor $h=h(r_1,\,r_2)=(1+|r_2|)(1-|r_1|)/(1-|r_1||r_2|)$ in Eq.~\eqref{eq:QGTI1}. Here, we consider the case of a perfectly reflecting bottom mirror, i.e. $h=h(r_1,\,1)=2$ which corresponds to the so-called imbalanced Gires–Tournois interferometer regime~\cite{GT-CRA-64}. These interferometers are widely used as optical pulse shaping elements~\cite{GAS-OL-00}. They induce a controllable amount and sign of group delay dispersion (GDD), which is typically the dominating effect outside resonance. The amount and sign of GDD are tunable by choosing the frequency of operation with respect to the cavity resonance and around resonance where the second-order contribution vanishes and switches sign, third order dispersion (TOD) becomes the leading term~\cite{SCM-PRL-19}. Due to TOD, the resulting temporal pulses can possess oscillatory tails, see Fig.~\ref{fig:fig1}~(b), where a typical periodic time trace obtained from integrating numerically Eqs.~\eqref{eq:QGTI1}-\eqref{eq:QGTI3}  is shown. As a result, the envelope of the corresponding OFC is asymmetrical, see Fig.~\ref{fig:fig1}~(c).

Finally, $\chi=\chi(u,D)$ is a nonlinear susceptibility function. As a simple model, we employ the quantum well susceptibility as introduced in Ref.~\cite{B-PRA-98} as
\begin{align}
        \chi=\frac{1}{\pi}\left[\ln(u+i-b)-2\ln(u+i-D)+\ln(u+i)-\ln(b)\right].\label{eq:QGTI_chi}
\end{align}
The susceptibility $\chi$ describes the response of the medium to incoming light with frequency $\omega_0$. This analytical formula originates from assuming two parabolic bands with a quasi-equilibrium Fermi distribution for electron and holes.
This response possesses both a real and an imaginary part that represents the non-instantaneous Kerr effect mediated by the population inversion, as well as the field saturable absorption, which also depends on $D$.
\begin{figure}
	\includegraphics[width=1\columnwidth]{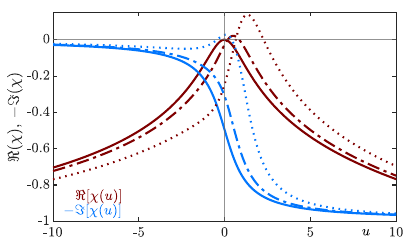}
	\caption{The real (red) and negative imaginary (blue) parts of the 
	susceptibility $\chi\left(u,D\right)$ as given by Eq.~\eqref{eq:QGTI_chi} for $b=100$ for $D=0.3$ (dash-dotted line) and $D=1.08$ (dotted line) which enclose the range of $D$ for the parameter set used in Fig. 1 as well as $D=0$ (solid line). \label{fig:fig2}}
	\end{figure}
Further, $\chi$ depends on $u$, the normalized frequency with respect to the band-gap frequency $\omega_G$ as $u=(\omega_0-\omega_G)/\gamma_{\perp}$, where $\gamma_{\perp}$ is the intraband polarization dephasing rate. The parameter $b$ relates to the top-band frequency $\omega_T$, and reads $b=(\omega_T-\omega_G)/\gamma_{\perp}$. The gain (or the saturable absorption) is defined as $-\Im(\chi)$ while the phase shift caused by the nonlinear refraction index is described by $\Re(\chi)$.

The situation where $D=0$ describes an empty conduction band and a transition from transparency to saturable absorption located around $u=0$ while $D=1$ corresponds to the optical transparency being reached at the bandgap, i.e. $\Im[\chi(0,1)]=0$. The real and negative imaginary parts of the susceptibility $\chi$ are represented for different values of the population inversion in Fig.~\ref{fig:fig2}.

\section{Bifurcation analysis of the DAE system}\label{sec:DAE}

\begin{figure}
	\includegraphics[width=1\columnwidth]{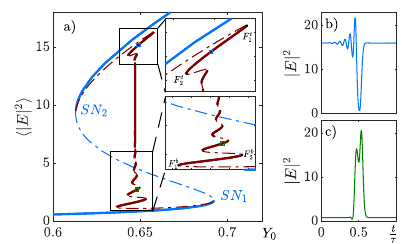}
		        \caption{(a) Branches of CW (blue) and TLSs (red) solutions of Eqs.~\eqref{eq:QGTI1}-\eqref{eq:QGTI3} as a function of the injection $Y_0$. Thick (thin) lines denote stable (unstable) solutions, respectively. The insets show zooms around the upper and lower snaking regions. In the right panels two exemplary dark (b) and bright (c) solution profiles at positions labeled with a blue and green square in (a) are displayed. Parameters as in Fig.~\ref{fig:fig1}~(b,c).\label{fig:fig3}}
	  \end{figure}

To understand the origin and the behaviour of the periodic pulse solution presented in Fig.~\ref{fig:fig1}~(b), we perform a bifurcation analysis of the system~\eqref{eq:QGTI1}-\eqref{eq:QGTI3} using a recently developed extension of DDE-BIFTOOL~\cite{DDEBT} that allows for the bifurcation analysis of algebraic and
neutral delayed equations~\cite{HGJ-OL-21,SJG-OL-22,KSG-OL-22,SGJ-PRL-22,BKW_JDEA_06}. We start the analysis with the situation where the injection is set to be resonant with an external cavity mode, i.e., $\varphi=0$. In the long-delay limit, where the time delay is the longest relevant time scale and in conjunction with sufficiently large detuning $\delta$, the quantum well nonlinearity causes a bistable CW response of the system in a certain parameter range of the injection intensity. We show in Fig.~\ref{fig:fig3}~(a) the resulting hysteresis diagram depicted in blue for the same parameter set used
in Figs.~\ref{fig:fig1}~(b,c). 
%
%
%
The stability is indicated by the line thickness. One can see that the CW branch loses stability in a saddle-node (SN) bifurcation at the point $SN_1$ and regains it in another SN bifurcation at $SN_2$. From the CW branch, one observes that the branch of the TLSs emerges in a subcritical Adronov–Hopf bifurcation. Indeed, in the bistable CW region, similar to the case of the injected microcavity subjected to the Kerr nonlinearity~\cite{SPV-OL-19,SJG-OL-22}, the coexistence of two stable CW solutions can lead to the formation of domain walls between them that exhibit oscillatory tails, for certain sets of parameters. In particular, two opposed fronts may lock at discrete distances forming stable bound states which results in the formation of TLSs. The corresponding branch of periodic solutions (cf. red line in Fig.~\ref{fig:fig3}~(a)) features stable dark and bright TLSs over an extended injection range. Here, different stable and unstable locking ranges occur along a collapsed snaking structure~\cite{KW_PD_05,PKG_PRA_16} around a Maxwell line where opposing fronts have the same drifting speed, see the insets in panel (a). On the left or the right of the Maxwell line, the drifting speeds differ, leading to a transient towards the lower or the upper CW state, respectively. Figure~\ref{fig:fig3}~(b,c) show two exemplary dark and bright solution profiles at different positions along the snaking branch. Remarkably, one notices that although being calculated for very close values of $Y_0$, the dark TLS shows a stronger oscillatory tail, while the snaking is more pronounced in the vicinity of the lower CW state, i.e., for the bright solutions. A video showing the whole evolution of the TLS solution along the branch is presented in Supplementary Material.
%
%
	  
\begin{figure}
	\includegraphics[width=1\columnwidth]{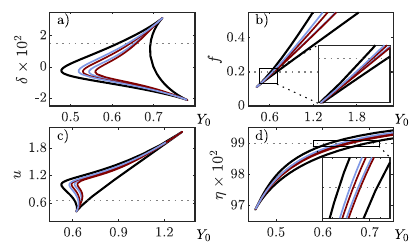}
		        \caption{(a)-(d) Two-parameter bifurcation diagrams in  $(Y_0,\,\delta)$, $(Y_0,\,f)$, $(Y_0,\,u)$ and $(Y_0,\,\eta)$ planes, respectively. Evolution of the points $SN_1$ and $SN_2$ limiting the region of CW bistability (black) is shown together with the two major folds limiting the stability of the dark (dark red) and bright (light blue) TLSs, respectively. The horizontal dashed lines indicate the particular parameter values used in Fig.~\ref{fig:fig3}. Other parameters as in Figs.~\ref{fig:fig1},~\ref{fig:fig3}.\label{fig:fig4}}
	  \end{figure}

To investigate the existence and the stability borders of the obtained TLSs solutions in this parameter regime, four two-parameter continuations in $(Y_0,\,\delta)$, $(Y_0,\,f)$, $(Y_0,\,u)$ and $(Y_0,\,\eta)$ planes are carried out and the results are presented in Fig.~\ref{fig:fig4} (a)-(d), respectively.
Here, black lines correspond to the thresholds of the SN bifurcation points $SN_1$ and $SN_2$ limiting the region of CW bistability, whereas dark red and light blue lines show the evolution of the two major folds limiting the stability of the dark and bright TLSs, respectively, cf. $F_{1,2}^{t,b}$ in the insets of Fig.~\ref{fig:fig3}~(a). The horizontal dashed line indicate the particular parameter values used in Fig.~\ref{fig:fig3}.     

In particular, in Fig.~\ref{fig:fig4}~(a) one can see that the CW bistabile region exists for both anomalous and normal dispersion regimes and the stability regions of both bright and dark TLSs is broader for small detunings and injection values. The latter vanish for large values of $\delta$ and $Y_0$. A similar trend is presented in panel (c), where the diagram in the $(Y_0,\, u)$ plane is shown. Here, one observes that the bistability region opens only in a certain range around $u\sim 1$, which corresponds to the most efficient nonlinear modulation of the losses. Again, the region of stable existence for TLSs decreases for increasing injection values. Figure~\ref{fig:fig4}~(b) presents the evolution of the corresponding folds in the effective gain prefactor $f$. Here, the bistability region starts to exist for small finite values of $f$ and small detuning values and it is becoming broader with increasing $f$, hence for stronger nonlinearty; the fold lines of both dark and bright TLSs follow the same trend. In panel (d), the dynamics of the limiting folds is shown in the $(Y_0,\,\eta)$ plane. Hence, the TLSs  solutions exist and are stable only in the so-called good cavity limit, where $\eta\rightarrow 1$. Moreover, the corresponding regions are moving in the direction of larger injections with increasing $\eta$. Note that this analysis was carried out for $\gamma=1$, i.e. assuming that the carrier lifetime is equal to the photon lifetime in the cavity, a situation that typically corresponds to a fast saturable absorber with a relaxation rate in the picosecond range.
%

\begin{figure}
	\includegraphics[width=1\columnwidth]{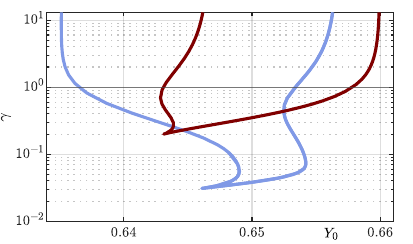}
		        \caption{Two-parameter bifurcation diagram in the $(Y_0,\,\gamma)$ plane showing the evolution of the major folds of the dark (dark red) and bright (light blue) TLSs. The horizontal black line indicates the particular parameter value used in Fig.~\ref{fig:fig3}. Other parameters as in Figs.~\ref{fig:fig1},~\ref{fig:fig3}\label{fig:fig5}.}
	  \end{figure}

Finally, we present in Fig.~\ref{fig:fig5} a continuation in the carrier lifetime $\gamma$. Here, the evolution of the major folds limiting the stability of the dark (dark red) and bright (light blue) TLSs is shown, cf. the inset in Fig.~\ref{fig:fig3}~(a). One can see that for increasing $\gamma$ values, first bright TLSs start to exist and the bistability between dark and bright TLSs is established only above some finite value of $\gamma$. Moreover, the width of the stability regions remains constant for large $\gamma$ values, where the adiabatic elimination of the carriers can be performed, connecting our results with past studies based upon an instantaneous nonlinearity~\cite{SPV-OL-19,SJG-OL-22,KSG-OL-22}.

\section{Dispersive master equation}\label{sec:PDE}
Although the bifurcation analysis allows for quantitative predictions of the behavior of the DAE system~\eqref{eq:QGTI1}-\eqref{eq:QGTI3}, an intuitive interpretation of the dynamics presented in Fig.~\ref{fig:fig3} is difficult since the role and impact of e.g., GDD and TOD on the TLS formation is hidden. Hence, to obtain a better understanding of the underlying physical mechanisms, we derive a partial differential equation (PDE) that approximates the dynamics of the full DAE model~\eqref{eq:QGTI1}-\eqref{eq:QGTI3} in the limit of a good cavity ($\eta\rightarrow 1$) and small detunings $\delta$ using a rigorous multiple time scale analysis as discussed in~\cite{SCM-PRL-19,SHJ-PRAp-20,SGJ-PRL-22,SJG-OL-22}.
We start our analysis by eliminating the field $Y(t)$  and transforming the DAE to a neutral delay differential equation (NDDE) that reads
\begin{eqnarray}
\eta\dot{E}(t-\tau)&+&\dot{E}(t)=\eta\left[if\chi(D(t-\tau))-1-i\delta+h\right]E(t-\tau)\nonumber\\
&+&\left[if\chi(D(t))-1-i\delta\right]E(t)+hY_0\sqrt{1-\eta^2}\,. \label{eq:NDDE}
\end{eqnarray}
Further, rescaling time with respect to the time-delay $\tau$ reveals a natural smallness parameter $\epsilon=\tau^{-1}$, where $\epsilon\ll 1$ in the long-delay limit. In the next step rescalings are chosen as  $E\rightarrow\epsilon^\frac{1}{2}\mathcal{E}$, $Y\rightarrow\epsilon^\frac{1}{2}\mathcal{Y}$, $Y_0\rightarrow\epsilon^\frac{1}{2}\mathcal{Y}_0$, and $D\rightarrow\epsilon\mathcal{D}$.  This choice of the scaling is also reflected in the treatment of the nonlinearity which now becomes $\chi(\epsilon\mathcal{D})$. Hence, the obvious next step is an expansion of $\chi$, leading to
\begin{equation}
 \chi(\epsilon\mathcal{D})=\chi(0)+\epsilon\chi^\prime(0)\mathcal{D}+\mathcal{O}(\epsilon^2)\,,
\end{equation}
where we ommited the $u$ dependence for brievity and
\begin{align}
\chi(0)=\chi_0&=-\frac{1}{\pi}\ln(-u-i)\nonumber\\
&=\frac{-1}{\pi}\ln\sqrt{u^2+1}+i\left[\frac{1}{2}+\frac{1}{\pi}\arctan(u)\right]\,,\\
\chi^\prime(0)=\chi_0^\prime&=\frac{2}{\pi}\frac{u-i}{u^2+1}\,.\nonumber
\end{align}
The dynamics is now considered on multiple timescales in powers of $\epsilon$.
 To this aim, the remaining fields and parameters are all expanded in powers of $\epsilon$. In addition, the fastest timescale $\sigma=\omega t$ is also stretched by expanding $\omega$ in powers of $\epsilon$ in order to account for the fact that the period $T$ is slightly larger than the time delay $\tau$. Hence, $\sigma$ describes the fast dynamics within one period.
 \begin{figure}
\includegraphics[width=1\columnwidth]{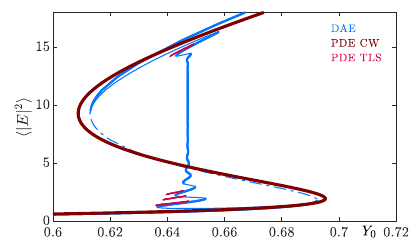}
\caption{\label{fig:test1} CW and exemplary numerical TLS solutions of the PDE system~\eqref{eq:PDE1}-\eqref{eq:PDE2} plotted together with the branches from Fig.~\ref{fig:fig3} of the DAE \eqref{eq:QGTI1}-\eqref{eq:QGTI3}. Also, some parameter values have been adjusted for the PDE results, since bistability is lost for the exact set of parameters that has been used  in Figs.~\ref{fig:fig1}, \ref{fig:fig3}. The adjusted parameters are $u=0.52$, $\delta=0.044$ and $\gamma=10$. The corresponding PDE parameters are shown in Table~\ref{table}.\label{fig:fig6}}
\end{figure}

After inserting all expansions into the NDDE~\eqref{eq:NDDE}, one collects orders of $\epsilon$ and solves them separately imposing the proper solvability conditions, see Supplementary Material for details. Finally, we find that the dispersive master PDE for the field $E$ on a fast timescale $\sigma$ containing the dynamics within one round-trip and an effective slow timescale $\theta$ that describes the dynamics over many round-trips, reads
\begin{equation}
 \partial_\theta E = \left(-i\,d_2\,\partial_\sigma^2+d_3\,\partial_\sigma^3\right)E+\left(\mathcal{L}+\mathcal{N}D\right)\,E+h\,Y_0\frac{\sqrt{1-\eta^2}}{1+i\delta}\,.\label{eq:PDE1}
\end{equation}
Here, the values of the coefficients $d_2$ and $d_3$ of the linear spatial
operator in Eq.~\eqref{eq:PDE1} are particularly instructive and read
\begin{equation}
 d_2=\frac{2\,\delta}{(1+\delta^2)^2}\,,\qquad d_3=\frac{2}{3}\frac{3\,\delta^2-1}{(1+\delta^2)^3}\,,
\end{equation}
whereas the coefficients $\mathcal{L}$ and $\mathcal{N}$ are
\begin{align}
    \mathcal{L}=&\,\eta-1-2\,\eta\,\arctan(\delta)\,i+2\,f\,\left(\frac{1-i\delta}{1+\delta^2}\right)^2\chi_0\,i,  \nonumber\\
    \mathcal{N}=&\,2\,f\,\frac{1-i\delta}{1+\delta^2}\,\chi^\prime_0\,i\,, \label{eq:PDE_coef}
\end{align}
Further, the equation for the carrier density $D$ takes the form
\begin{equation}
 \gamma^{-1}\,\partial_\sigma D=-D+\Im{\left(\chi_0+\chi^\prime_0D\right)}|E|^2\,.\label{eq:PDE2}
\end{equation}
One can see that the second order dispersion coefficient $d_2$ in Eq.~\eqref{eq:PDE1} cancels at resonance, i.e., at $\delta=0$, corresponding to the transition from anomalous to normal dispersion regime. Hence, at small $\delta$ values the dynamics is mostly controlled by TOD, which explains the oscillating tails of the TLSs observed in Fig.~\ref{fig:fig3}. Further, TOD vanishes for $\delta=\delta_c=\pm1/\sqrt{3}$.

Note that the linear  operator in Eq.~\eqref{eq:PDE1}  corresponding to the case of the linear empty cavity (i.e., $D=0$ (and $\chi=0$)) coincides with the linear part of the normal form PDE derived for the case of a quasi-conservative dispersive microcavity containing a Kerr medium coupled to a distant external mirror~\cite{SGJ-PRL-22}. There, a similar multiscale analysis performed in the long delay limit for small cavity losses and weak injection yielded the Lugiato-Lefever equation with TOD~\cite{LL-PRL-87,PGL-OL-14,PGG-PRA-17}. The latter can be understood as a dissipative version of the nonlinear Schr\"odinger equation in which the phase symmetry is broken by the presence of monochromatic injection with amplitude $Y_0$ and frequency offset $\delta$ with respect to the microcavity resonance. In the presence of the nonlinear element, the corresponding  contributions depending on $\chi_0$, $\chi'_{0}$, $f$ as well as $\delta$  appear in the coefficients $\mathcal{L}$, $\mathcal{N}$, cf. Eqs.~\eqref{eq:PDE_coef}.
\begin{figure}
\includegraphics[width=1\columnwidth]{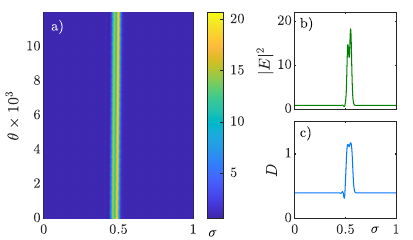}
\caption{\label{fig:fig7} (a): Space-time evolution of a bright TLS obtained from the numerical simulations of the PDE model~\eqref{eq:PDE1}-\eqref{eq:PDE2}  for  $Y_0=0.6437$ (b),(c): Exemplary profiles taken at the last time-step showing the intensity  $|E|^2$ and $D$, respectively.}
\end{figure}

The CW solution $(E_s,\, D(E_s))$ of Eqs.~\eqref{eq:PDE1}-\eqref{eq:PDE2} can be found analytically by solving them for $Y_0 $ yielding
\begin{align}
    Y_0^2&=\frac{1+\delta^2}{h^2(1-\eta^2)}|\mathcal{L}+\mathcal{N}D(I_s)|^2\,I_s\,, \label{eq:CW1}\\
    D(I_s)&=\frac{\Im{(\chi_0)}\,I_s}{1-\Im{(\chi_0^\prime)}\,I_s}\,,\label{eq:CW2}
\end{align}
where $I_s=|E_s|^2$.

Notice that the CW states \eqref{eq:CW1}-\eqref{eq:CW2} does not exhibit a bistability region for the same parameter set as the CW solution obtained from the DAE system parameters corresponding to the solution presented in Fig.~\ref{fig:fig3}. However, for a slightly different set of system parameters where the parameters that were scaled or expanded during the PDE~\eqref{eq:PDE1}-\eqref{eq:PDE2} derivation are adjusted, the bistability region  is established  and almost identical to the results obtained from the  DAE system ~\eqref{eq:QGTI1}-\eqref{eq:QGTI3}, see Fig.~\ref{fig:fig6}. Here, the DAE CW and TLS solutions (blue) are plotted together with the corresponding PDE CW (dark red) and exemplary TLS (light red) solutions, obtained by means of direct numerical simulations. The corresponding PDE coefficients are presented in Table~\ref{table}.

\begin{figure}
\includegraphics[width=1\columnwidth]{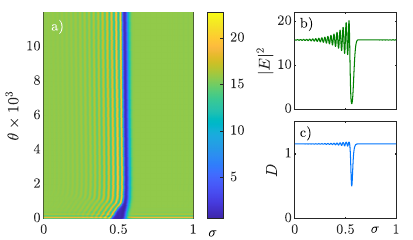}
\caption{\label{fig:fig8} (a): Space-time evolution of a dark TLS obtained from the numerical simulations of the PDE model~\eqref{eq:PDE1}-\eqref{eq:PDE2}  for $Y_0=0.6491$. (b),(c): Exemplary profiles taken at the last time-step showing the intensitiy  $|E|^2$ and $D$, respectively.}
\end{figure}

The observed discrepancy can be ascribed to the fact that assumption of a small population inversion $D\ll 1$ is not always strictly verified, see for instance Fig.~\ref{fig:fig7}~(c) and Fig.~\ref{fig:fig8}~(c) where $D$ is of order $\mathcal{O}(1)$ during TLS emission, but also on the CW background. 
\begin{table}[ht]

	\centering
\begin{tabular}{ |p{1cm}|p{2.5cm}|  }
	\hline
	$d_2$ & $0.088$  \\
	$d_3$ & $-0.659$    \\
	$\mathcal{L}$ & $-0.271 - 0.079i$ \\
	$\mathcal{N}$    & $0.204 + 0.095i$ \\
	$\chi_0$ & $-0.038 + 0.653i$   \\
	$\chi_0^\prime$ & $0.261 - 0.501i$ \\
	\hline
\end{tabular}
\caption{Coefficients for the PDE \eqref{eq:PDE1}-\eqref{eq:PDE2} solutions presented in Fig.~\ref{fig:fig6}. Other parameters are as in Fig.~\ref{fig:fig3}}
\label{table}
\end{table}

In order to compare the solutions of the normal form  \eqref{eq:PDE1}-\eqref{eq:PDE2}  with
those obtained from the full DAE model~\eqref{eq:QGTI1}-\eqref{eq:QGTI3}, we solved the PDE master equations  \eqref{eq:PDE1}-\eqref{eq:PDE2} numerically for the adjusted parameter set from Fig.~\ref{fig:fig1} and Fig.~\ref{fig:fig3}. The direct numerical simulations were performed using a standard split-step based pseudo-spectral method~\cite{GJ-PRA-17}. The carrier equation~\eqref{eq:PDE2} was integrated using a dynamical boundary condition as in~\cite{Hausen20}.

To visualize the evolution of the resulting solution over many round-trips, a space-time representation in $(\sigma,\,\theta)$ plane is employed and the results are presented in Figs. ~\ref{fig:fig7},~\ref{fig:fig8}. Figure \ref{fig:fig7}~(a) shows the time evolution of a bright TLS.  The corresponding profiles at the end of the simulations are presented in panels (b) and (c) for the intensity $|E|^2$ and carrier density $D$, respectively. Here, one can see that  the domain walls lock at the second locking distance and the corresponding profile agrees with the bright TLS found in the DAE system, cf. Fig~\ref{fig:fig3}~(c).
An example of a  dark TLS is shown in Fig.~\ref{fig:fig8}~(a). The initial condition evolves into a dark TLS with strong asymmetric oscillatory tail as seen in an exemplary profiles in (b) and (c) for $E$ and $D$, respectively. Again, this solution is in good agreement with the dark TLS found in the original DAE model, see Fig~\ref{fig:fig3}~(b).
Note that both dark TLSs from the PDE and DAE possess more oscillatory tails as the bright TLSs.  This situation is similar to the one observed for the Lugiato-Lefever equation subjected to TOD in the normal dispersion regime~\cite{PGG-PRA-17}. There, TOD allows for stable dark and bright TLSs to coexist, while the shape of the switching waves connecting the top and bottom CW solutions is modified and bright TLSs are stabilized due to the generation of additional oscillations in the switching-wave profiles.

\section{Square-Waves}\label{sec:SW}
Finally, we consider the same system given by DAE~\eqref{eq:QGTI1}-\eqref{eq:QGTI3}  but operated out of the bistability regime such that the wavelength of the optical injection is set in-between two modes of the external cavity. This situation corresponds to the case of antiresonant feedback, i.e. $\varphi=\pi$.

Note that the formation of SWs was reported in various optical and opto-electronic systems subjected to time-delayed feedback~\cite{PJC-PRE-09,JAH-PRL-15,MGJ-PRA-07,GES-OL-06,LLZ-OL-16}.  There, they typically appear in the long delay limit  via a supercritical Andronov-Hopf (AH) bifurcation as modulated oscillations with a period close to twice the value of the time-delay $\tau$. The SWs then evolve into sharp transition layers connecting plateaus of duration $\sim\tau$. Recently, the formation of SWs in a monomode micro-cavity containing a Kerr medium coupled to a long external feedback cavity under CW  injection in the normal dispersion regime was demonstrated~\cite{KSG-OL-22}. There, SWs also appear via a homoclinic snaking scenario, leading to the formation of complex-shaped multistable SW solutions.
 \begin{figure}
	\includegraphics[width=1\columnwidth]{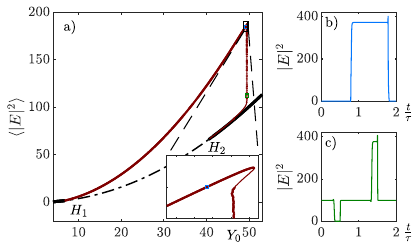}
	\caption{SWs emerging in a supercritical AH bifurcation at $H_1$ and a subcritical AH bifurcation at $H_2$. Such a branch will be referred to as a subcritical SW branch of the QGTI. b) shows a stable SW with $50\%$ duty cycle, while c) shows a solution with three plateaus at the Maxwell line.  $(\delta,h,\eta,\varphi,\gamma,f,u,b,\tau)=(-0.01,2,0.98,\pi,0.1,1,0.65,1000,1000)$. \label{fig:fig9}}
	\end{figure}
To study the formation and the properties of the SWs in the DAE model~\eqref{eq:QGTI1}-\eqref{eq:QGTI3},  we perform the path-continuation as a function of the injection $Y_0$,  setting the feedback phase $\varphi=\pi$. Our results are shown in Fig.~\ref{fig:fig9}. In Fig.~\ref{fig:fig9}~(a), for increasing $Y_0$, a branch of stable SW solutions (dark red) emerges from the stable CW state (black) at a supercritical AH bifurcation at the point $H_1$. Here, a small-amplitude  periodic solution with a period $\sim 2\tau$ close to the bifurcation point transforms into a well-developed SW with sharp transition layers  when $Y_0$ is increased, see Fig.~\ref{fig:fig9}~(b), where the exemplary profile is shown (cf. the blue point in the panel (a)). However, for further increased $Y_0$, the SW branch turns in a saddle-node bifurcation (cf. the inset in the panel (a)) and experiences a sequence of further saddle-node bifurcations  resulting in a collapsed snaking structure. In particular, close to  the vertical Maxwell line where the plateaus shrink in favor of the CW value, it leads to a three-plateau solution as shown in panel (c). Finally, the SW branch merges with the CW state at $H_2$ in a subcritical AH bifurcation. A video showing the whole evolution of the SW solution along the branch is presented in Supplementary Material. Remarkably, the snaking scenario is similar to one reported in ~\cite{KSG-OL-22}, where, however, it was observed for low injection values, whereas in the present case the subcritical AH bifurcation is created at high $Y_0$.

\begin{figure}
	\includegraphics[width=1\columnwidth]{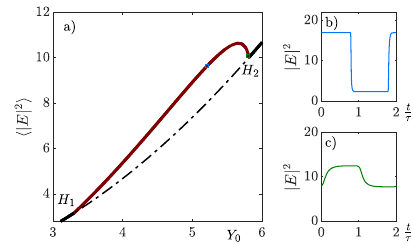}
	\caption{An exemplary branch for the supercritical regime, close to the transition from sub- to supercritical SW solutions. The two profiles in b) and c) show a fully developed as well as a just emerging SW, respectively. The parameters are $(\delta,h,\eta,\varphi,\gamma,f,u,b,\tau)=(-0.01,2,0.86,\pi,0.1,1,0.65,1000,1000)$ \label{fig:fig10}.}
	\end{figure}

By varying $\eta$ to lower values one can observe a transition of the AH bifurcation at $H_2$ from subcritical to supercritical, which leads to a disappearance of the snaking structure. The transition seems to be located close to the SWs existing threshold in $\eta$. A resulting branch can be seen in Fig.~ \ref{fig:fig10}~(a). While panel (b) demonstrates an exemplary profile of a  fully developed SW with sharp transition layers, panel (c) shows a smoother periodic solution close to the (now) supercritical AH bifurcation at $H_2$.
	
\section*{Conclusion}

In  conclusion, in this paper  the dynamics of temporal localized states and square-waves in semiconductor micro-resonators with strong time-delayed feedback was analyzed. We considered a semiconductor quantum-well as the nonlinear element instead of an ideal, instantaneous Kerr response. This allowed us to study the influence of the band-gap detuning with respect to the injection field and the population inversion time scale onto the dynamics. A first principle model for the optical response was used in combination with time delayed algebraic constraints that correspond to the boundary conditions linking the optical field in the various parts of the compound cavity system.

Using a combination of analytical, numerical and path-continuation methods, the existence of  dark and bright temporal localized states  appearing via the locking of domain walls that connect the high and low bistable continuous wave  intensity levels was demonstrated in the normal dispersion regime and for resonant feedback. Further, operated in the antiresonant (pi-shifted) feedback regime, square-waves solutions possessing a period of approximately twice the round-trip in the external cavity were obtained. Depending on the reflectivity of the external mirror, the  latter can also appear through a homoclinic snaking scenario. Finally, to clarify the role of the second and third order dispersion in the formation of the localized states, a multiple time-scale analysis in the good cavity limit was performed. The resulting dispersive normal form partial differential equation shows a good qualitative agreement with the original time-delayed model.

More generally, the results presented here may serve as a guide for further analysis into the dynamics of  temporal localized states in semiconductor micro-resonators subjected to the strong time-delayed feedback in the long cavity regime. Our analysis clarified that, even in the good cavity limit, when the carrier lifetime becomes much smaller than the micro-cavity photon lifetime, both branches of bright and dark TLSs disappear, even if CW optical bistability is not affected. This indicates that the most favorable situation would certainly consider a quantum-well medium optimized to act as a saturable absorber with a fast carrier recovery time in the picosecond range instead of, e.g., a micro-cavity containing a gain medium engineered to favor a lasing action and for which the carrier lifetime would be in the nanosecond range.

\section*{Supplementary Material}
See supplementary material for the animation showing the solution profiles along the branches shown in Figs. \ref{fig:fig3} and \ref{fig:fig9} in Sec.~\ref{sec:DAE} and Sec.~\ref{sec:SW}, respecitely and for details on the derivation of the normal form partial differential equation presented in Sec.~\ref{sec:PDE}.
\section*{Author Declarations}
The authors have no conflicts to disclose.
\section*{Acknowledgments}
T.S. thanks the foundation “Studienstiftung des deutschen Volkes” for financial support,  E.K, J. J. and S.G. acknowledge the financial support of the projects KEFIR, Ministerio de Economía y Competitividad (PID2021-128910NB-100 AEI/FEDER UE) and KOGIT, Agence Nationale de la Recherche (ANR-22-CE92-0009)), Deutsche Forschungsgemeinschaft (DFG) via Grant Nr. 505936983.
We thank A. Giacomotti and B. Garbin for useful discussions.
\section*{Data Availability}
The data that support the findings of this study are available from the corresponding author upon reasonable request.

\section*{References}

\begin{thebibliography}{35}%
\makeatletter
\providecommand \@ifxundefined [1]{%
 \@ifx{#1\undefined}
}%
\providecommand \@ifnum [1]{%
 \ifnum #1\expandafter \@firstoftwo
 \else \expandafter \@secondoftwo
 \fi
}%
\providecommand \@ifx [1]{%
 \ifx #1\expandafter \@firstoftwo
 \else \expandafter \@secondoftwo
 \fi
}%
\providecommand \natexlab [1]{#1}%
\providecommand \enquote  [1]{``#1''}%
\providecommand \bibnamefont  [1]{#1}%
\providecommand \bibfnamefont [1]{#1}%
\providecommand \citenamefont [1]{#1}%
\providecommand \href@noop [0]{\@secondoftwo}%
\providecommand \href [0]{\begingroup \@sanitize@url \@href}%
\providecommand \@href[1]{\@@startlink{#1}\@@href}%
\providecommand \@@href[1]{\endgroup#1\@@endlink}%
\providecommand \@sanitize@url [0]{\catcode `\\12\catcode `\$12\catcode
  `\&12\catcode `\#12\catcode `\^12\catcode `\_12\catcode `\%12\relax}%
\providecommand \@@startlink[1]{}%
\providecommand \@@endlink[0]{}%
\providecommand \url  [0]{\begingroup\@sanitize@url \@url }%
\providecommand \@url [1]{\endgroup\@href {#1}{\urlprefix }}%
\providecommand \urlprefix  [0]{URL }%
\providecommand \Eprint [0]{\href }%
\providecommand \doibase [0]{https://doi.org/}%
\providecommand \selectlanguage [0]{\@gobble}%
\providecommand \bibinfo  [0]{\@secondoftwo}%
\providecommand \bibfield  [0]{\@secondoftwo}%
\providecommand \translation [1]{[#1]}%
\providecommand \BibitemOpen [0]{}%
\providecommand \bibitemStop [0]{}%
\providecommand \bibitemNoStop [0]{.\EOS\space}%
\providecommand \EOS [0]{\spacefactor3000\relax}%
\providecommand \BibitemShut  [1]{\csname bibitem#1\endcsname}%
\let\auto@bib@innerbib\@empty
\bibitem [{\citenamefont {Cundiff}\ and\ \citenamefont {Ye}(2003)}]{CY-RMP-03}%
  \BibitemOpen
  \bibfield  {author} {\bibinfo {author} {\bibfnamefont {S.~T.}\ \bibnamefont
  {Cundiff}}\ and\ \bibinfo {author} {\bibfnamefont {J.}~\bibnamefont {Ye}},\
  }\href {https://doi.org/10.1103/RevModPhys.75.325} {\bibfield  {journal}
  {\bibinfo  {journal} {Rev. Mod. Phys.}\ }\textbf {\bibinfo {volume} {75}},\
  \bibinfo {pages} {325} (\bibinfo {year} {2003})}\BibitemShut {NoStop}%
\bibitem [{\citenamefont {Udem}\ \emph {et~al.}(2002)\citenamefont {Udem},
  \citenamefont {Holzwarth},\ and\ \citenamefont {H{\"a}nsch}}]{UHH-NAT-02}%
  \BibitemOpen
  \bibfield  {author} {\bibinfo {author} {\bibfnamefont {T.}~\bibnamefont
  {Udem}}, \bibinfo {author} {\bibfnamefont {R.}~\bibnamefont {Holzwarth}},\
  and\ \bibinfo {author} {\bibfnamefont {T.~W.}\ \bibnamefont {H{\"a}nsch}},\
  }\href {https://doi.org/10.1038/416233a} {\bibfield  {journal} {\bibinfo
  {journal} {Nature}\ }\textbf {\bibinfo {volume} {416}},\ \bibinfo {pages}
  {233} (\bibinfo {year} {2002})}\BibitemShut {NoStop}%
\bibitem [{\citenamefont {Cundiff}\ and\ \citenamefont
  {Weiner}(2010)}]{CW-NAP-10}%
  \BibitemOpen
  \bibfield  {author} {\bibinfo {author} {\bibfnamefont {S.~T.}\ \bibnamefont
  {Cundiff}}\ and\ \bibinfo {author} {\bibfnamefont {A.~M.}\ \bibnamefont
  {Weiner}},\ }\href {https://doi.org/10.1038/nphoton.2010.196} {\bibfield
  {journal} {\bibinfo  {journal} {Nature Photonics}\ }\textbf {\bibinfo
  {volume} {4}},\ \bibinfo {pages} {760} (\bibinfo {year} {2010})}\BibitemShut
  {NoStop}%
\bibitem [{\citenamefont {Diddams}(2010)}]{D-JOSAB-10}%
  \BibitemOpen
  \bibfield  {author} {\bibinfo {author} {\bibfnamefont {S.~A.}\ \bibnamefont
  {Diddams}},\ }\href {https://doi.org/10.1364/JOSAB.27.000B51} {\bibfield
  {journal} {\bibinfo  {journal} {J. Opt. Soc. Am. B}\ }\textbf {\bibinfo
  {volume} {27}},\ \bibinfo {pages} {B51} (\bibinfo {year} {2010})}\BibitemShut
  {NoStop}%
\bibitem [{\citenamefont {Pasquazi}\ \emph {et~al.}(2018)\citenamefont
  {Pasquazi}, \citenamefont {Peccianti}, \citenamefont {Razzari}, \citenamefont
  {Moss}, \citenamefont {Coen}, \citenamefont {Erkintalo}, \citenamefont
  {Chembo}, \citenamefont {Hansson}, \citenamefont {Wabnitz}, \citenamefont
  {Del'Haye}, \citenamefont {Xue}, \citenamefont {Weiner},\ and\ \citenamefont
  {Morandotti}}]{PPR-PR-18}%
  \BibitemOpen
  \bibfield  {author} {\bibinfo {author} {\bibfnamefont {A.}~\bibnamefont
  {Pasquazi}}, \bibinfo {author} {\bibfnamefont {M.}~\bibnamefont {Peccianti}},
  \bibinfo {author} {\bibfnamefont {L.}~\bibnamefont {Razzari}}, \bibinfo
  {author} {\bibfnamefont {D.~J.}\ \bibnamefont {Moss}}, \bibinfo {author}
  {\bibfnamefont {S.}~\bibnamefont {Coen}}, \bibinfo {author} {\bibfnamefont
  {M.}~\bibnamefont {Erkintalo}}, \bibinfo {author} {\bibfnamefont {Y.~K.}\
  \bibnamefont {Chembo}}, \bibinfo {author} {\bibfnamefont {T.}~\bibnamefont
  {Hansson}}, \bibinfo {author} {\bibfnamefont {S.}~\bibnamefont {Wabnitz}},
  \bibinfo {author} {\bibfnamefont {P.}~\bibnamefont {Del'Haye}}, \bibinfo
  {author} {\bibfnamefont {X.}~\bibnamefont {Xue}}, \bibinfo {author}
  {\bibfnamefont {A.~M.}\ \bibnamefont {Weiner}},\ and\ \bibinfo {author}
  {\bibfnamefont {R.}~\bibnamefont {Morandotti}},\ }\href
  {https://doi.org/https://doi.org/10.1016/j.physrep.2017.08.004} {\bibfield
  {journal} {\bibinfo  {journal} {Physics Reports}\ }\textbf {\bibinfo {volume}
  {729}},\ \bibinfo {pages} {1 } (\bibinfo {year} {2018})}\BibitemShut
  {NoStop}%
\bibitem [{\citenamefont {Herr}\ \emph {et~al.}(2014)\citenamefont {Herr},
  \citenamefont {Brasch}, \citenamefont {Jost}, \citenamefont {Wang},
  \citenamefont {Kondratiev}, \citenamefont {Gorodetsky},\ and\ \citenamefont
  {Kippenberg}}]{HBJ-NAP-14}%
  \BibitemOpen
  \bibfield  {author} {\bibinfo {author} {\bibfnamefont {T.}~\bibnamefont
  {Herr}}, \bibinfo {author} {\bibfnamefont {V.}~\bibnamefont {Brasch}},
  \bibinfo {author} {\bibfnamefont {J.~D.}\ \bibnamefont {Jost}}, \bibinfo
  {author} {\bibfnamefont {C.~Y.}\ \bibnamefont {Wang}}, \bibinfo {author}
  {\bibfnamefont {N.~M.}\ \bibnamefont {Kondratiev}}, \bibinfo {author}
  {\bibfnamefont {M.~L.}\ \bibnamefont {Gorodetsky}},\ and\ \bibinfo {author}
  {\bibfnamefont {T.~J.}\ \bibnamefont {Kippenberg}},\ }\href
  {https://doi.org/10.1038/nphoton.2013.343} {\bibfield  {journal} {\bibinfo
  {journal} {Nature Photonics}\ }\textbf {\bibinfo {volume} {8}},\ \bibinfo
  {pages} {145} (\bibinfo {year} {2014})}\BibitemShut {NoStop}%
\bibitem [{\citenamefont {Leo}\ \emph {et~al.}(2010)\citenamefont {Leo},
  \citenamefont {Coen}, \citenamefont {Kockaert}, \citenamefont {Gorza},
  \citenamefont {Emplit},\ and\ \citenamefont {Haelterman}}]{LCK-NAP-10}%
  \BibitemOpen
  \bibfield  {author} {\bibinfo {author} {\bibfnamefont {F.}~\bibnamefont
  {Leo}}, \bibinfo {author} {\bibfnamefont {S.}~\bibnamefont {Coen}}, \bibinfo
  {author} {\bibfnamefont {P.}~\bibnamefont {Kockaert}}, \bibinfo {author}
  {\bibfnamefont {S.}~\bibnamefont {Gorza}}, \bibinfo {author} {\bibfnamefont
  {P.}~\bibnamefont {Emplit}},\ and\ \bibinfo {author} {\bibfnamefont
  {M.}~\bibnamefont {Haelterman}},\ }\href
  {https://doi.org/10.1038/nphoton.2010.120} {\bibfield  {journal} {\bibinfo
  {journal} {Nat Photon}\ }\textbf {\bibinfo {volume} {4}},\ \bibinfo {pages}
  {471} (\bibinfo {year} {2010})}\BibitemShut {NoStop}%
\bibitem [{\citenamefont {Tropper}\ \emph {et~al.}(2004)\citenamefont
  {Tropper}, \citenamefont {Foreman}, \citenamefont {Garnache}, \citenamefont
  {Wilcox},\ and\ \citenamefont {Hoogland}}]{tropper04}%
  \BibitemOpen
  \bibfield  {author} {\bibinfo {author} {\bibfnamefont {A.~C.}\ \bibnamefont
  {Tropper}}, \bibinfo {author} {\bibfnamefont {H.~D.}\ \bibnamefont
  {Foreman}}, \bibinfo {author} {\bibfnamefont {A.}~\bibnamefont {Garnache}},
  \bibinfo {author} {\bibfnamefont {K.~G.}\ \bibnamefont {Wilcox}},\ and\
  \bibinfo {author} {\bibfnamefont {S.~H.}\ \bibnamefont {Hoogland}},\
  }\href@noop {} {\bibfield  {journal} {\bibinfo  {journal} {J. Phys. D: Appl.
  Phys.}\ }\textbf {\bibinfo {volume} {37}},\ \bibinfo {pages} {R75} (\bibinfo
  {year} {2004})}\BibitemShut {NoStop}%
\bibitem [{\citenamefont {Laurain}\ \emph {et~al.}(2010)\citenamefont
  {Laurain}, \citenamefont {Myara}, \citenamefont {Beaudoin}, \citenamefont
  {Sagnes},\ and\ \citenamefont {Garnache}}]{LMB-OE-10}%
  \BibitemOpen
  \bibfield  {author} {\bibinfo {author} {\bibfnamefont {A.}~\bibnamefont
  {Laurain}}, \bibinfo {author} {\bibfnamefont {M.}~\bibnamefont {Myara}},
  \bibinfo {author} {\bibfnamefont {G.}~\bibnamefont {Beaudoin}}, \bibinfo
  {author} {\bibfnamefont {I.}~\bibnamefont {Sagnes}},\ and\ \bibinfo {author}
  {\bibfnamefont {A.}~\bibnamefont {Garnache}},\ }\href
  {https://doi.org/10.1364/OE.18.014627} {\bibfield  {journal} {\bibinfo
  {journal} {Opt. Express}\ }\textbf {\bibinfo {volume} {18}},\ \bibinfo
  {pages} {14627} (\bibinfo {year} {2010})}\BibitemShut {NoStop}%
\bibitem [{\citenamefont {Wang}\ \emph {et~al.}(2017)\citenamefont {Wang},
  \citenamefont {Van~Gasse}, \citenamefont {Moskalenko}, \citenamefont
  {Latkowski}, \citenamefont {Bente}, \citenamefont {Kuyken},\ and\
  \citenamefont {Roelkens}}]{WVM-LSP-17}%
  \BibitemOpen
  \bibfield  {author} {\bibinfo {author} {\bibfnamefont {Z.}~\bibnamefont
  {Wang}}, \bibinfo {author} {\bibfnamefont {K.}~\bibnamefont {Van~Gasse}},
  \bibinfo {author} {\bibfnamefont {V.}~\bibnamefont {Moskalenko}}, \bibinfo
  {author} {\bibfnamefont {S.}~\bibnamefont {Latkowski}}, \bibinfo {author}
  {\bibfnamefont {E.}~\bibnamefont {Bente}}, \bibinfo {author} {\bibfnamefont
  {B.}~\bibnamefont {Kuyken}},\ and\ \bibinfo {author} {\bibfnamefont
  {G.}~\bibnamefont {Roelkens}},\ }\href
  {http://dx.doi.org/10.1038/lsa.2016.260} {\bibfield  {journal} {\bibinfo
  {journal} {Light: Science $\&$ Applications}\ }\textbf {\bibinfo {volume}
  {6}},\ \bibinfo {pages} {e16260 EP } (\bibinfo {year} {2017})},\ \bibinfo
  {note} {original Article}\BibitemShut {NoStop}%
\bibitem [{\citenamefont {Schelte}\ \emph
  {et~al.}(2019{\natexlab{a}})\citenamefont {Schelte}, \citenamefont {Pimenov},
  \citenamefont {Vladimirov}, \citenamefont {Javaloyes},\ and\ \citenamefont
  {Gurevich}}]{SPV-OL-19}%
  \BibitemOpen
  \bibfield  {author} {\bibinfo {author} {\bibfnamefont {C.}~\bibnamefont
  {Schelte}}, \bibinfo {author} {\bibfnamefont {A.}~\bibnamefont {Pimenov}},
  \bibinfo {author} {\bibfnamefont {A.~G.}\ \bibnamefont {Vladimirov}},
  \bibinfo {author} {\bibfnamefont {J.}~\bibnamefont {Javaloyes}},\ and\
  \bibinfo {author} {\bibfnamefont {S.~V.}\ \bibnamefont {Gurevich}},\ }\href
  {https://doi.org/10.1364/OL.44.004925} {\bibfield  {journal} {\bibinfo
  {journal} {Opt. Lett.}\ }\textbf {\bibinfo {volume} {44}},\ \bibinfo {pages}
  {4925} (\bibinfo {year} {2019}{\natexlab{a}})}\BibitemShut {NoStop}%
\bibitem [{\citenamefont {Seidel}\ \emph
  {et~al.}(2022{\natexlab{a}})\citenamefont {Seidel}, \citenamefont
  {Javaloyes},\ and\ \citenamefont {Gurevich}}]{SJG-OL-22}%
  \BibitemOpen
  \bibfield  {author} {\bibinfo {author} {\bibfnamefont {T.~G.}\ \bibnamefont
  {Seidel}}, \bibinfo {author} {\bibfnamefont {J.}~\bibnamefont {Javaloyes}},\
  and\ \bibinfo {author} {\bibfnamefont {S.~V.}\ \bibnamefont {Gurevich}},\
  }\href {https://doi.org/10.1364/OL.457777} {\bibfield  {journal} {\bibinfo
  {journal} {Opt. Lett.}\ }\textbf {\bibinfo {volume} {47}},\ \bibinfo {pages}
  {2979} (\bibinfo {year} {2022}{\natexlab{a}})}\BibitemShut {NoStop}%
\bibitem [{\citenamefont {Koch}\ \emph {et~al.}(2022)\citenamefont {Koch},
  \citenamefont {Seidel}, \citenamefont {Gurevich},\ and\ \citenamefont
  {Javaloyes}}]{KSG-OL-22}%
  \BibitemOpen
  \bibfield  {author} {\bibinfo {author} {\bibfnamefont {E.~R.}\ \bibnamefont
  {Koch}}, \bibinfo {author} {\bibfnamefont {T.~G.}\ \bibnamefont {Seidel}},
  \bibinfo {author} {\bibfnamefont {S.~V.}\ \bibnamefont {Gurevich}},\ and\
  \bibinfo {author} {\bibfnamefont {J.}~\bibnamefont {Javaloyes}},\ }\href
  {https://doi.org/10.1364/OL.468236} {\bibfield  {journal} {\bibinfo
  {journal} {Opt. Lett.}\ }\textbf {\bibinfo {volume} {47}},\ \bibinfo {pages}
  {4343} (\bibinfo {year} {2022})}\BibitemShut {NoStop}%
\bibitem [{\citenamefont {Schelte}\ \emph
  {et~al.}(2019{\natexlab{b}})\citenamefont {Schelte}, \citenamefont {Camelin},
  \citenamefont {Marconi}, \citenamefont {Garnache}, \citenamefont {Huyet},
  \citenamefont {Beaudoin}, \citenamefont {Sagnes}, \citenamefont {Giudici},
  \citenamefont {Javaloyes},\ and\ \citenamefont {Gurevich}}]{SCM-PRL-19}%
  \BibitemOpen
  \bibfield  {author} {\bibinfo {author} {\bibfnamefont {C.}~\bibnamefont
  {Schelte}}, \bibinfo {author} {\bibfnamefont {P.}~\bibnamefont {Camelin}},
  \bibinfo {author} {\bibfnamefont {M.}~\bibnamefont {Marconi}}, \bibinfo
  {author} {\bibfnamefont {A.}~\bibnamefont {Garnache}}, \bibinfo {author}
  {\bibfnamefont {G.}~\bibnamefont {Huyet}}, \bibinfo {author} {\bibfnamefont
  {G.}~\bibnamefont {Beaudoin}}, \bibinfo {author} {\bibfnamefont
  {I.}~\bibnamefont {Sagnes}}, \bibinfo {author} {\bibfnamefont
  {M.}~\bibnamefont {Giudici}}, \bibinfo {author} {\bibfnamefont
  {J.}~\bibnamefont {Javaloyes}},\ and\ \bibinfo {author} {\bibfnamefont
  {S.~V.}\ \bibnamefont {Gurevich}},\ }\href
  {https://doi.org/10.1103/PhysRevLett.123.043902} {\bibfield  {journal}
  {\bibinfo  {journal} {Phys. Rev. Lett.}\ }\textbf {\bibinfo {volume} {123}},\
  \bibinfo {pages} {043902} (\bibinfo {year} {2019}{\natexlab{b}})}\BibitemShut
  {NoStop}%
\bibitem [{\citenamefont {Mulet}\ and\ \citenamefont
  {Balle}(2005)}]{MB-JQE-05}%
  \BibitemOpen
  \bibfield  {author} {\bibinfo {author} {\bibfnamefont {J.}~\bibnamefont
  {Mulet}}\ and\ \bibinfo {author} {\bibfnamefont {S.}~\bibnamefont {Balle}},\
  }\href {https://doi.org/10.1109/JQE.2005.853355} {\bibfield  {journal}
  {\bibinfo  {journal} {Quantum Electronics, IEEE Journal of}\ }\textbf
  {\bibinfo {volume} {41}},\ \bibinfo {pages} {1148} (\bibinfo {year}
  {2005})}\BibitemShut {NoStop}%
\bibitem [{\citenamefont {Gires}\ and\ \citenamefont
  {Tournois}(1964)}]{GT-CRA-64}%
  \BibitemOpen
  \bibfield  {author} {\bibinfo {author} {\bibfnamefont {F.}~\bibnamefont
  {Gires}}\ and\ \bibinfo {author} {\bibfnamefont {P.}~\bibnamefont
  {Tournois}},\ }\href@noop {} {\bibfield  {journal} {\bibinfo  {journal} {C.
  R. Acad. Sci. Paris}\ }\textbf {\bibinfo {volume} {258}},\ \bibinfo {pages}
  {6112} (\bibinfo {year} {1964})}\BibitemShut {NoStop}%
\bibitem [{\citenamefont {Golubovic}\ \emph {et~al.}(2000)\citenamefont
  {Golubovic}, \citenamefont {Austin}, \citenamefont {Steiner-Shepard},
  \citenamefont {Reed}, \citenamefont {Diddams}, \citenamefont {Jones},\ and\
  \citenamefont {Engen}}]{GAS-OL-00}%
  \BibitemOpen
  \bibfield  {author} {\bibinfo {author} {\bibfnamefont {B.}~\bibnamefont
  {Golubovic}}, \bibinfo {author} {\bibfnamefont {R.~R.}\ \bibnamefont
  {Austin}}, \bibinfo {author} {\bibfnamefont {M.~K.}\ \bibnamefont
  {Steiner-Shepard}}, \bibinfo {author} {\bibfnamefont {M.~K.}\ \bibnamefont
  {Reed}}, \bibinfo {author} {\bibfnamefont {S.~A.}\ \bibnamefont {Diddams}},
  \bibinfo {author} {\bibfnamefont {D.~J.}\ \bibnamefont {Jones}},\ and\
  \bibinfo {author} {\bibfnamefont {A.~G.~V.}\ \bibnamefont {Engen}},\ }\href
  {https://doi.org/10.1364/OL.25.000275} {\bibfield  {journal} {\bibinfo
  {journal} {Opt. Lett.}\ }\textbf {\bibinfo {volume} {25}},\ \bibinfo {pages}
  {275} (\bibinfo {year} {2000})}\BibitemShut {NoStop}%
\bibitem [{\citenamefont {Balle}(1998)}]{B-PRA-98}%
  \BibitemOpen
  \bibfield  {author} {\bibinfo {author} {\bibfnamefont {S.}~\bibnamefont
  {Balle}},\ }\href@noop {} {\bibfield  {journal} {\bibinfo  {journal} {Phys.
  Rev. A}\ }\textbf {\bibinfo {volume} {57}},\ \bibinfo {pages} {1304}
  (\bibinfo {year} {1998})}\BibitemShut {NoStop}%
\bibitem [{\citenamefont {Engelborghs}\ \emph {et~al.}(2002)\citenamefont
  {Engelborghs}, \citenamefont {Luzyanina},\ and\ \citenamefont
  {Roose}}]{DDEBT}%
  \BibitemOpen
  \bibfield  {author} {\bibinfo {author} {\bibfnamefont {K.}~\bibnamefont
  {Engelborghs}}, \bibinfo {author} {\bibfnamefont {T.}~\bibnamefont
  {Luzyanina}},\ and\ \bibinfo {author} {\bibfnamefont {D.}~\bibnamefont
  {Roose}},\ }\href {https://doi.org/10.1145/513001.513002} {\bibfield
  {journal} {\bibinfo  {journal} {ACM Trans. Math. Softw.}\ }\textbf {\bibinfo
  {volume} {28}},\ \bibinfo {pages} {1} (\bibinfo {year} {2002})}\BibitemShut
  {NoStop}%
\bibitem [{\citenamefont {Hessel}\ \emph {et~al.}(2021)\citenamefont {Hessel},
  \citenamefont {Gurevich},\ and\ \citenamefont {Javaloyes}}]{HGJ-OL-21}%
  \BibitemOpen
  \bibfield  {author} {\bibinfo {author} {\bibfnamefont {D.}~\bibnamefont
  {Hessel}}, \bibinfo {author} {\bibfnamefont {S.~V.}\ \bibnamefont
  {Gurevich}},\ and\ \bibinfo {author} {\bibfnamefont {J.}~\bibnamefont
  {Javaloyes}},\ }\href {https://doi.org/10.1364/OL.428182} {\bibfield
  {journal} {\bibinfo  {journal} {Opt. Lett.}\ }\textbf {\bibinfo {volume}
  {46}},\ \bibinfo {pages} {2557} (\bibinfo {year} {2021})}\BibitemShut
  {NoStop}%
\bibitem [{\citenamefont {Seidel}\ \emph
  {et~al.}(2022{\natexlab{b}})\citenamefont {Seidel}, \citenamefont
  {Gurevich},\ and\ \citenamefont {Javaloyes}}]{SGJ-PRL-22}%
  \BibitemOpen
  \bibfield  {author} {\bibinfo {author} {\bibfnamefont {T.~G.}\ \bibnamefont
  {Seidel}}, \bibinfo {author} {\bibfnamefont {S.~V.}\ \bibnamefont
  {Gurevich}},\ and\ \bibinfo {author} {\bibfnamefont {J.}~\bibnamefont
  {Javaloyes}},\ }\href {https://doi.org/10.1103/PhysRevLett.128.083901}
  {\bibfield  {journal} {\bibinfo  {journal} {Phys. Rev. Lett.}\ }\textbf
  {\bibinfo {volume} {128}},\ \bibinfo {pages} {083901} (\bibinfo {year}
  {2022}{\natexlab{b}})}\BibitemShut {NoStop}%
\bibitem [{\citenamefont {Barton}\ \emph {et~al.}(2006)\citenamefont {Barton},
  \citenamefont {Krauskopf},\ and\ \citenamefont {Wilson}}]{BKW_JDEA_06}%
  \BibitemOpen
  \bibfield  {author} {\bibinfo {author} {\bibfnamefont {D.~A.~W.}\
  \bibnamefont {Barton}}, \bibinfo {author} {\bibfnamefont {B.}~\bibnamefont
  {Krauskopf}},\ and\ \bibinfo {author} {\bibfnamefont {R.~E.}\ \bibnamefont
  {Wilson}},\ }\href {https://doi.org/10.1080/10236190601045663} {\bibfield
  {journal} {\bibinfo  {journal} {Journal of Difference Equations and
  Applications}\ }\textbf {\bibinfo {volume} {12}},\ \bibinfo {pages} {1087}
  (\bibinfo {year} {2006})},\ \Eprint
  {https://arxiv.org/abs/https://doi.org/10.1080/10236190601045663}
  {https://doi.org/10.1080/10236190601045663} \BibitemShut {NoStop}%
\bibitem [{\citenamefont {Knobloch}\ and\ \citenamefont
  {Wagenknecht}(2005)}]{KW_PD_05}%
  \BibitemOpen
  \bibfield  {author} {\bibinfo {author} {\bibfnamefont {J.}~\bibnamefont
  {Knobloch}}\ and\ \bibinfo {author} {\bibfnamefont {T.}~\bibnamefont
  {Wagenknecht}},\ }\href
  {https://doi.org/https://doi.org/10.1016/j.physd.2005.04.018} {\bibfield
  {journal} {\bibinfo  {journal} {Physica D: Nonlinear Phenomena}\ }\textbf
  {\bibinfo {volume} {206}},\ \bibinfo {pages} {82 } (\bibinfo {year}
  {2005})}\BibitemShut {NoStop}%
\bibitem [{\citenamefont {Parra-Rivas}\ \emph {et~al.}(2016)\citenamefont
  {Parra-Rivas}, \citenamefont {Knobloch}, \citenamefont {Gomila},\ and\
  \citenamefont {Gelens}}]{PKG_PRA_16}%
  \BibitemOpen
  \bibfield  {author} {\bibinfo {author} {\bibfnamefont {P.}~\bibnamefont
  {Parra-Rivas}}, \bibinfo {author} {\bibfnamefont {E.}~\bibnamefont
  {Knobloch}}, \bibinfo {author} {\bibfnamefont {D.}~\bibnamefont {Gomila}},\
  and\ \bibinfo {author} {\bibfnamefont {L.}~\bibnamefont {Gelens}},\ }\href
  {https://doi.org/10.1103/PhysRevA.93.063839} {\bibfield  {journal} {\bibinfo
  {journal} {Phys. Rev. A}\ }\textbf {\bibinfo {volume} {93}},\ \bibinfo
  {pages} {063839} (\bibinfo {year} {2016})}\BibitemShut {NoStop}%
\bibitem [{\citenamefont {Schelte}\ \emph {et~al.}(2020)\citenamefont
  {Schelte}, \citenamefont {Hessel}, \citenamefont {Javaloyes},\ and\
  \citenamefont {Gurevich}}]{SHJ-PRAp-20}%
  \BibitemOpen
  \bibfield  {author} {\bibinfo {author} {\bibfnamefont {C.}~\bibnamefont
  {Schelte}}, \bibinfo {author} {\bibfnamefont {D.}~\bibnamefont {Hessel}},
  \bibinfo {author} {\bibfnamefont {J.}~\bibnamefont {Javaloyes}},\ and\
  \bibinfo {author} {\bibfnamefont {S.~V.}\ \bibnamefont {Gurevich}},\ }\href
  {https://doi.org/10.1103/PhysRevApplied.13.054050} {\bibfield  {journal}
  {\bibinfo  {journal} {Phys. Rev. Applied}\ }\textbf {\bibinfo {volume}
  {13}},\ \bibinfo {pages} {054050} (\bibinfo {year} {2020})}\BibitemShut
  {NoStop}%
\bibitem [{\citenamefont {Lugiato}\ and\ \citenamefont
  {Lefever}(1987)}]{LL-PRL-87}%
  \BibitemOpen
  \bibfield  {author} {\bibinfo {author} {\bibfnamefont {L.~A.}\ \bibnamefont
  {Lugiato}}\ and\ \bibinfo {author} {\bibfnamefont {R.}~\bibnamefont
  {Lefever}},\ }\href {https://doi.org/10.1103/PhysRevLett.58.2209} {\bibfield
  {journal} {\bibinfo  {journal} {Phys. Rev. Lett.}\ }\textbf {\bibinfo
  {volume} {58}},\ \bibinfo {pages} {2209} (\bibinfo {year}
  {1987})}\BibitemShut {NoStop}%
\bibitem [{\citenamefont {Parra-Rivas}\ \emph {et~al.}(2014)\citenamefont
  {Parra-Rivas}, \citenamefont {Gomila}, \citenamefont {Leo}, \citenamefont
  {Coen},\ and\ \citenamefont {Gelens}}]{PGL-OL-14}%
  \BibitemOpen
  \bibfield  {author} {\bibinfo {author} {\bibfnamefont {P.}~\bibnamefont
  {Parra-Rivas}}, \bibinfo {author} {\bibfnamefont {D.}~\bibnamefont {Gomila}},
  \bibinfo {author} {\bibfnamefont {F.}~\bibnamefont {Leo}}, \bibinfo {author}
  {\bibfnamefont {S.}~\bibnamefont {Coen}},\ and\ \bibinfo {author}
  {\bibfnamefont {L.}~\bibnamefont {Gelens}},\ }\href
  {https://doi.org/10.1364/OL.39.002971} {\bibfield  {journal} {\bibinfo
  {journal} {Opt. Lett.}\ }\textbf {\bibinfo {volume} {39}},\ \bibinfo {pages}
  {2971} (\bibinfo {year} {2014})}\BibitemShut {NoStop}%
\bibitem [{\citenamefont {Parra-Rivas}\ \emph {et~al.}(2017)\citenamefont
  {Parra-Rivas}, \citenamefont {Gomila},\ and\ \citenamefont
  {Gelens}}]{PGG-PRA-17}%
  \BibitemOpen
  \bibfield  {author} {\bibinfo {author} {\bibfnamefont {P.}~\bibnamefont
  {Parra-Rivas}}, \bibinfo {author} {\bibfnamefont {D.}~\bibnamefont
  {Gomila}},\ and\ \bibinfo {author} {\bibfnamefont {L.}~\bibnamefont
  {Gelens}},\ }\href {https://doi.org/10.1103/PhysRevA.95.053863} {\bibfield
  {journal} {\bibinfo  {journal} {Phys. Rev. A}\ }\textbf {\bibinfo {volume}
  {95}},\ \bibinfo {pages} {053863} (\bibinfo {year} {2017})}\BibitemShut
  {NoStop}%
\bibitem [{\citenamefont {Gurevich}\ and\ \citenamefont
  {Javaloyes}(2017)}]{GJ-PRA-17}%
  \BibitemOpen
  \bibfield  {author} {\bibinfo {author} {\bibfnamefont {S.~V.}\ \bibnamefont
  {Gurevich}}\ and\ \bibinfo {author} {\bibfnamefont {J.}~\bibnamefont
  {Javaloyes}},\ }\href {https://doi.org/10.1103/PhysRevA.96.023821} {\bibfield
   {journal} {\bibinfo  {journal} {Phys. Rev. A}\ }\textbf {\bibinfo {volume}
  {96}},\ \bibinfo {pages} {023821} (\bibinfo {year} {2017})}\BibitemShut
  {NoStop}%
\bibitem [{\citenamefont {Hausen}\ \emph {et~al.}(2020)\citenamefont {Hausen},
  \citenamefont {L\"{u}dge}, \citenamefont {Gurevich},\ and\ \citenamefont
  {Javaloyes}}]{Hausen20}%
  \BibitemOpen
  \bibfield  {author} {\bibinfo {author} {\bibfnamefont {J.}~\bibnamefont
  {Hausen}}, \bibinfo {author} {\bibfnamefont {K.}~\bibnamefont {L\"{u}dge}},
  \bibinfo {author} {\bibfnamefont {S.~V.}\ \bibnamefont {Gurevich}},\ and\
  \bibinfo {author} {\bibfnamefont {J.}~\bibnamefont {Javaloyes}},\ }\href
  {https://doi.org/10.1364/OL.406136} {\bibfield  {journal} {\bibinfo
  {journal} {Opt. Lett.}\ }\textbf {\bibinfo {volume} {45}},\ \bibinfo {pages}
  {6210} (\bibinfo {year} {2020})}\BibitemShut {NoStop}%
\bibitem [{\citenamefont {Peil}\ \emph {et~al.}(2009)\citenamefont {Peil},
  \citenamefont {Jacquot}, \citenamefont {Chembo}, \citenamefont {Larger},\
  and\ \citenamefont {Erneux}}]{PJC-PRE-09}%
  \BibitemOpen
  \bibfield  {author} {\bibinfo {author} {\bibfnamefont {M.}~\bibnamefont
  {Peil}}, \bibinfo {author} {\bibfnamefont {M.}~\bibnamefont {Jacquot}},
  \bibinfo {author} {\bibfnamefont {Y.~K.}\ \bibnamefont {Chembo}}, \bibinfo
  {author} {\bibfnamefont {L.}~\bibnamefont {Larger}},\ and\ \bibinfo {author}
  {\bibfnamefont {T.}~\bibnamefont {Erneux}},\ }\href
  {https://doi.org/10.1103/PhysRevE.79.026208} {\bibfield  {journal} {\bibinfo
  {journal} {Phys. Rev. E}\ }\textbf {\bibinfo {volume} {79}},\ \bibinfo
  {pages} {026208} (\bibinfo {year} {2009})}\BibitemShut {NoStop}%
\bibitem [{\citenamefont {Javaloyes}\ \emph {et~al.}(2015)\citenamefont
  {Javaloyes}, \citenamefont {Ackemann},\ and\ \citenamefont
  {Hurtado}}]{JAH-PRL-15}%
  \BibitemOpen
  \bibfield  {author} {\bibinfo {author} {\bibfnamefont {J.}~\bibnamefont
  {Javaloyes}}, \bibinfo {author} {\bibfnamefont {T.}~\bibnamefont
  {Ackemann}},\ and\ \bibinfo {author} {\bibfnamefont {A.}~\bibnamefont
  {Hurtado}},\ }\href {https://doi.org/10.1103/PhysRevLett.112.223901}
  {\bibfield  {journal} {\bibinfo  {journal} {Phys. Rev. Lett.}\ }\textbf
  {\bibinfo {volume} {115}},\ \bibinfo {pages} {223901} (\bibinfo {year}
  {2015})}\BibitemShut {NoStop}%
\bibitem [{\citenamefont {Mulet}\ \emph {et~al.}(2007)\citenamefont {Mulet},
  \citenamefont {Giudici}, \citenamefont {Javaloyes},\ and\ \citenamefont
  {Balle}}]{MGJ-PRA-07}%
  \BibitemOpen
  \bibfield  {author} {\bibinfo {author} {\bibfnamefont {J.}~\bibnamefont
  {Mulet}}, \bibinfo {author} {\bibfnamefont {M.}~\bibnamefont {Giudici}},
  \bibinfo {author} {\bibfnamefont {J.}~\bibnamefont {Javaloyes}},\ and\
  \bibinfo {author} {\bibfnamefont {S.}~\bibnamefont {Balle}},\ }\href
  {https://doi.org/10.1103/PhysRevA.76.043801} {\bibfield  {journal} {\bibinfo
  {journal} {Phys. Rev. A}\ }\textbf {\bibinfo {volume} {76}},\ \bibinfo
  {pages} {043801} (\bibinfo {year} {2007})}\BibitemShut {NoStop}%
\bibitem [{\citenamefont {Gavrielides}\ \emph {et~al.}(2006)\citenamefont
  {Gavrielides}, \citenamefont {Erneux}, \citenamefont {Sukow}, \citenamefont
  {Burner}, \citenamefont {McLachlan}, \citenamefont {Miller},\ and\
  \citenamefont {Amonette}}]{GES-OL-06}%
  \BibitemOpen
  \bibfield  {author} {\bibinfo {author} {\bibfnamefont {A.}~\bibnamefont
  {Gavrielides}}, \bibinfo {author} {\bibfnamefont {T.}~\bibnamefont {Erneux}},
  \bibinfo {author} {\bibfnamefont {D.~W.}\ \bibnamefont {Sukow}}, \bibinfo
  {author} {\bibfnamefont {G.}~\bibnamefont {Burner}}, \bibinfo {author}
  {\bibfnamefont {T.}~\bibnamefont {McLachlan}}, \bibinfo {author}
  {\bibfnamefont {J.}~\bibnamefont {Miller}},\ and\ \bibinfo {author}
  {\bibfnamefont {J.}~\bibnamefont {Amonette}},\ }\href
  {https://doi.org/10.1364/OL.31.002006} {\bibfield  {journal} {\bibinfo
  {journal} {Opt. Lett.}\ }\textbf {\bibinfo {volume} {31}},\ \bibinfo {pages}
  {2006} (\bibinfo {year} {2006})}\BibitemShut {NoStop}%
\bibitem [{\citenamefont {Li}\ \emph {et~al.}(2016)\citenamefont {Li},
  \citenamefont {Li}, \citenamefont {Zhuang}, \citenamefont {Mezosi},
  \citenamefont {Sorel},\ and\ \citenamefont {Chan}}]{LLZ-OL-16}%
  \BibitemOpen
  \bibfield  {author} {\bibinfo {author} {\bibfnamefont {S.-S.}\ \bibnamefont
  {Li}}, \bibinfo {author} {\bibfnamefont {X.-Z.}\ \bibnamefont {Li}}, \bibinfo
  {author} {\bibfnamefont {J.-P.}\ \bibnamefont {Zhuang}}, \bibinfo {author}
  {\bibfnamefont {G.}~\bibnamefont {Mezosi}}, \bibinfo {author} {\bibfnamefont
  {M.}~\bibnamefont {Sorel}},\ and\ \bibinfo {author} {\bibfnamefont {S.-C.}\
  \bibnamefont {Chan}},\ }\href {https://doi.org/10.1364/OL.41.000812}
  {\bibfield  {journal} {\bibinfo  {journal} {Opt. Lett.}\ }\textbf {\bibinfo
  {volume} {41}},\ \bibinfo {pages} {812} (\bibinfo {year} {2016})}\BibitemShut
  {NoStop}%
\end{thebibliography}
%
%
\end{document}


\section*{Supplementary material: Normal Form derivation}

In this supplementary material we demonstrate how a partial differential equation (PDE) normal form model can be derived from our first principle Delay-Algebraic Equation (DAE) for a micro-cavity operating in the Gires-Tournois regime containing a quantum well and coupled to an long external cavity under CW injection using multiple time scaling. The resulting PDE system can help to get a better understanding on how the parameters influence the dynamics of temporal localized states in the underlying system. The derivation will be done in a good cavity limit, i.e. $\eta\rightarrow 1$ and $h=2$.

\section{Normal Form Settings}
We start from the full DAE system which reads
\begin{align}
    \dot{E}(t)&=\left[if\chi -1 -i\delta\right]E(t)+hY(t),\label{eq:QGTI_1}\\ 
    \dot{D}(t)&=\gamma\left[-D(t)+\Im(\chi)|E|^2\right],\label{eq:QGTI_2}\\
    Y(t)&=\eta e^{i\varphi}\left[E(t-\tau)-Y(t-\tau)\right]+Y_0\sqrt{1-\eta^2}, \label{eq:QGTI_3}
\end{align}
with
\begin{align}
    \chi(u,D)=\frac{1}{\pi}\left(\ln(u+i-b)-2\ln(u+i-D)+\ln(u+i)-\ln(b)\right). \label{eq:QGTI_chi}
\end{align}
Here, we only consider the case $\varphi=0$. Going from a DAE to a neutral delay differential equation (NDDE) allows to eliminate $Y(t)$. This can be achieved by adding Eq. \eqref{eq:QGTI_1} at time $t$ onto itself at time $t-\tau$, multiplied by $\eta$. One can then recognize and insert Eq. \eqref{eq:QGTI_3}, which eliminates $Y(t)$ and leads to
\begin{align}
    \left(\dot{E}(t)+\eta\dot{E}(t-\tau)\right)=&\left[if\chi(D(t))-1-i\delta\right]E(t)+\eta\left[if\chi(D(t-1))-1-i\delta\right]E(t-\tau)\nonumber\\&+h\eta E(t-\tau)+hY_0\sqrt(1-\eta^2).
\end{align}
\subsection{Time Scaling}
First, the system is rescaled with $t\rightarrow\frac{t}{\tau}$ and $\frac{1}{\tau}=\epsilon$, yielding
\begin{align}
    \epsilon\left(\dot{E}(t)+\eta\dot{E}(t-1)\right)=&\left[if\chi(D(t))-1-i\delta\right]E(t)+\eta\left[if\chi(D(t-1))-1-i\delta\right]E(t-1)\nonumber\\
    &+h\eta E(t-1)+hY_0\sqrt(1-\eta^2)\label{eq:rescaled_QGTI1_3},\\
   \epsilon\dot{D}=& \gamma\left[-D(t)+\Im(\chi(D))|E|^2\right].
\end{align}
It now has a period of $T \approx 1$ defining the fastest timescale $t_0$. To avoid drift, the period $T$ or equivalently the frequency $\omega$ can be expanded allowing for higher order corrections. This leads to a stretch in $t_0$:
\begin{align}
    t_0&=\omega t,\\
    t_1&=0,\\
    t_2&=\epsilon^2t,\\
    t_3&=\epsilon^3t.
\end{align}
\subsection{Parameter Expansions}
In the following, the parameters will be expanded based on the knowledge about the system as well as some practical thoughts:
\begin{align}
    Y_0&=\epsilon\gamma_1+\epsilon^2\gamma_2+\epsilon^3\gamma_3,\\
    \eta&=1+\eta^2\eta_2.
\end{align}
As mentioned before, the normal form is derived in a good cavity limit by choosing $\eta\rightarrow1$ and $h=2$ eliminating $h$ from the equation. Furthermore, choosing $Y_0$ being $\mathcal{O}(\epsilon)$ combined with the choice of $\eta$  leads to an injection that comes in at $\mathcal{O}(\epsilon^2)$. The detuning $\delta$ will not be expanded, nevertheless $\delta$ will be assumed to be small at some point of the derivation. The parameters $u$ and $b$ only appear within $\chi$, hence they will not be considered for the moment. 
Finally, the major parameter to be discussed is $f$ determining the strength of the nonlinearity.
\subsection{Nonlinearity}
In general the nonlinearity should enter at $\mathcal{O}(\epsilon^3)$ limiting nonlinear effects. The most obvious path is the choice $f=\mathcal{O}(\epsilon^3)$. In that case however, it appears to be hard to access the nonlinearity itself from the argument of $\chi$. Another way is to choose $f$ in a lower order  such as $f\rightarrow \epsilon^2 f$ and rescale the field, as the following example will demonstrate. Consider $E\rightarrow\epsilon^\frac{1}{2} \mathcal{E}$, $Y\rightarrow\epsilon^\frac{1}{2} \mathcal{Y}$ and $Y_0\rightarrow \epsilon^\frac{1}{2} \mathcal{Y}_0$. This leads to:
\begin{align}
    \epsilon \dot{D}(t)&=\gamma\left[-D(t)+\epsilon\Im(\chi(D))|\mathcal{E}|^2\right].
\end{align}
Now $D\rightarrow \epsilon \mathcal{D}$ is an obvious consequence to be able to divide each term by $\epsilon$ yielding
\begin{align}
     \epsilon\dot{\mathcal{D}}(t)&=\gamma\left[-\mathcal{D}(t)+\Im(\chi(\epsilon \mathcal{D}))|\mathcal{E}|^2\right],
\end{align}
as well as $\chi(\epsilon \mathcal{D})$ in general. For Eq. \eqref{eq:rescaled_QGTI1_3} one finds
\begin{align}
    \epsilon^\frac{1}{2}\epsilon\left(\dot{\mathcal{E}}(t)+\eta\dot{\mathcal{E}}(t-1)\right)=&\epsilon^\frac{1}{2}\left[i\epsilon^2f\chi(\epsilon \mathcal{D})-1-i\delta\right]\mathcal{E}(t)\nonumber\\
    &+\epsilon^\frac{1}{2}\eta\left[i\epsilon^2f\chi(\epsilon \mathcal{D}(t-1))-1-i\delta\right]\mathcal{E}(t-1)\nonumber\\
    &+\epsilon^\frac{1}{2}h\eta \mathcal{E}(t-1)+\epsilon^\frac{1}{2} h\mathcal{Y}_0\sqrt(1-\eta^2). \label{eq:insert1}
\end{align}
Here $\epsilon^\frac{1}{2}$ can be taken away from each term to reproduce the old equation only differing in the argument of $\chi$. To write the susceptibility as a polynomial in $\epsilon$, a Taylor expansion can be included for small $\epsilon$, leading to:
\begin{align}
    \chi(\epsilon \mathcal{D})=&\chi(0)+\epsilon \chi'(0)\mathcal{D} + \mathcal{O}(\epsilon^2).
\end{align}
Note that the higher order terms will not occur in the final equation if only a total order of $\mathcal{O}(\epsilon^3)$ is considered. The terms $\chi(0)$ and $\chi'(0)$ are the complex terms
\begin{align}
   \chi(0)&=\frac{1}{\pi}\ln\left(\frac{u+i-b}{(u+i)b}\right)=\frac{1}{\pi}\ln\left(\frac{1}{b}-\frac{1}{u+i}\right)\approx\frac{1}{\pi}\ln\left(\frac{-1}{u+i}\right)=-\frac{1}{\pi}\ln(-u-i),\\
   \chi'(0)&=\frac{1}{\pi}\frac{2}{u+i-0}=\frac{1}{\pi}\frac{2(u-i)}{u^2+1},
\end{align}
which do not depend on the fields. Here, $b\gg 1$ has been used. This approximation is optional and can be omitted easily, if needed, since the susceptibility will be inserted in the final equation. The imaginary part of $\chi(0)$ can be isolated using the Arg function:
\begin{align}
    -\ln(-u-i)=&-\ln|u+i|-i\text{Arg}(-u-i)\\
    =&-\ln\sqrt{u^2+1}+i\left[\frac{\pi}{2}+\arctan(u)\right].
\end{align}
This leads to
\begin{align}
    \chi(0)&=-\frac{1}{\pi}\ln\sqrt{u^2+1}+i\left[\frac{1}{2}+\frac{1}{\pi}\arctan(u)\right]=R_0+iI_0,\\
    \chi'(0)&=\frac{2}{\pi}\frac{u-i}{u^2+1}=R_1+iI_1,
\end{align}
which yields an approximated susceptibility 
\begin{align}
    \chi(\epsilon \mathcal{D})=R_0+iI_0+\epsilon(R_1+iI_1)\mathcal{D}+\mathcal{O}(\epsilon^2),
\end{align}
with the nonlinear term in $\mathcal{O}(\epsilon^3)$. Inserting these results one finds for $\dot{\mathcal{D}}$
\begin{align}
   \epsilon  \dot{\mathcal{D}}(t)&=\gamma\left[-\mathcal{D}(t)+\Im(\chi(\epsilon \mathcal{D}))|\mathcal{E}|^2\right]\\
     &=\gamma\left[-\mathcal{D}(t)+(I_0+\epsilon I_1 \mathcal{D}(t))|\mathcal{E}(t)|^2\right]. \label{eq:insert2}
\end{align}

\section{Derivation}
\subsection{Time Derivative and Drift Operator}
Now the corresponding time derivative operator is:
\begin{align}
    \partial_t=\omega\partial_0+\epsilon^2\partial_2+\epsilon^3\partial_3.
\end{align}
To equalize the drift the expression $\omega=1+\epsilon\omega_1+\epsilon^2\omega_2+\epsilon^2\omega_3$ can be used which leads to:
\begin{align}
    \partial_t&=\partial_0+\epsilon\omega_1\partial_0+\epsilon^2(\omega_2\partial_+\partial_2)+\epsilon^3(\omega_3\partial_0+\partial_3)+\mathcal{O}(\epsilon^4)\\
    &=\mathcal{T}_0+\epsilon\mathcal{T}_1+\epsilon^2\mathcal{T}_2+\epsilon^3\mathcal{T}_3.
\end{align}
The field has to be written with respect to the new time scale. Thus, $\mathcal{E}(t)=\mathcal{E}(t_0,t_2,t_3)$, but also $\mathcal{E}(t-1)=\mathcal{E}(t_0-\omega,t_2-\epsilon^2,t_3-\epsilon^3)$ has to be used. In order to sort the system by orders of $\epsilon$ an expansion for small $\epsilon$ can be used to rewrite the time-delayed field
\begin{align}
    \mathcal{E}(t_0-\omega,t_2-\epsilon^2,t_3-\epsilon^3)=&\mathcal{E}(t_0-1-\epsilon\omega_1-\epsilon^2\omega_2-\epsilon^3\omega_3,t_2-\epsilon^2,t_3-\epsilon^3)\\
    =&\mathcal{E}(t_0-1,t_2,t_3)+\epsilon(-\omega_1\partial_0)\mathcal{E}(t_0-1,t_2,t_3)\nonumber\\
    &+\epsilon^2\left(\frac{\omega_1^2\partial_0^2}{2}-\omega_2\partial_0-\partial_2\right)\mathcal{E}(t_0-1,t_2,t_3)\\\nonumber&+\epsilon^3\left(\frac{-\omega_1^3\partial_0^3}{6}+\omega_1\omega_2\partial_0^2+\omega_1\partial_0\partial_2-\omega_3\partial_0-\partial_3\right)\mathcal{E}(t_0-1,t_2,t_3)
\end{align}
The found prefactors correspond to the shift operator $\mathcal{S}$ fulfilling
\begin{align}
    \mathcal{E}(t_0-\omega,t_2-\epsilon^2,t_3-\epsilon^3)&=\mathcal{S}\mathcal{E}(t_0-1)\\
    &=\left(1+\epsilon\mathcal{S}_1+\epsilon^2\mathcal{S}_2+\epsilon^3\mathcal{S}_3\right)\mathcal{E}(t_0-1),
\end{align}
with the reduced notation $\mathcal{E}(t_0-1,t_2,t_3)=\mathcal{E}(t_0-1)$.\\
Using the operators introduced above all definitions can be inserted into Eq. \eqref{eq:insert1} and Eq. \eqref{eq:insert2}. Additionally, the fields are expanded in a power series as $\mathcal{E}=\mathcal{E}_0+\epsilon\mathcal{E}_1+\epsilon^2\mathcal{E}_2+\epsilon^3\mathcal{E}_3+\mathcal{O}(\epsilon^4)$. Afterwards the resulting expressions will be sorted in orders of $\epsilon$ which will be considered separately. 

\subsection{Solving by Orders}
Starting with the lowest order $\mathcal{O}(\epsilon^0)$ one obtains
\begin{align}
\mathcal{E}_0(t_0)-\mathcal{E}_0(t_0-1)=0,
\end{align}
describing a simple periodic solution. \\
Using this, from $\mathcal{O}(\epsilon^1)$
\begin{align}
\mathcal{E}_1(t_0)-\mathcal{E}_1(t_0-1)&=\left(-\omega_1-\frac{2}{\delta^2+1}\right)\partial_0\mathcal{E}_0(t_0),
\end{align}
is found.
This equation describes a periodic solution with a periodicity disturbed by a drift term elucidating the meaning of the expansion in $\omega$ as compensation for potential drift. Without the first order term $\omega_1\partial_0\mathcal{E}_0(t_0)$ a nonvanishing drift of $\mathcal{E}_0$ on the time scale $t_0$ is observed. The drift is compensated through the result for the small frequency correction $\epsilon\omega_1$ by choosing
\begin{align}
    \omega_1=\frac{-2}{1+\delta^2}.
\end{align}
For $\mathcal{O}(\epsilon^2)$, the lower orders will be used once again leading to $\omega_2=\omega_1^3$. Furthermore, the solvability condition
\begin{align}
    \partial_2\mathcal{E}_0=\left[-i\delta\frac{\omega_1^2}{2}\partial_0^2-L_2\right]\mathcal{E}_0+\Gamma_2,
\end{align}
is found with
\begin{align}
    L_2&=\frac{2(fI_0-ifR_0)}{(1+i\delta)^2}-(\eta_2+i\theta_2),\\
    \Gamma_2&=\gamma_1\frac{2\sqrt{-2\eta_2}}{1+i\delta}.
\end{align}
The operator $L_2$ contains the first term from the expansion of the susceptibility, but the nonlinearity itself does not enter until $\mathcal{O}(\epsilon^3)$. 
Here, $\omega_3=\omega_1^3$ is found.
The final solvability condition for the third order containing all simplifications, has the form
\begin{align}
    \partial_3\mathcal{E}_0+\partial_2\mathcal{E}_1=&-i\frac{3\delta}{2}\omega_1^3\partial_0^2\mathcal{E}_0+\frac{1-3\delta^2}{12}\omega_1^3\partial_0^3\mathcal{E}_0\nonumber\\
    &+[\mathcal{N}\mathcal{D}-\omega_1L_2]\mathcal{E}_0+\left[-i\delta\frac{\omega_1^2}{2}\partial_0^2-L_2\right]\mathcal{E}_1+\Gamma_3+\omega_1\Gamma_2,
\end{align}
with
\begin{align}
    L_2&=\frac{2(fI_0-ifR_0)}{(1+i\delta)^2}-(\eta_2+i\theta_2),\\
    \Gamma_2&=\gamma_1\frac{2\sqrt{-2\eta_2}}{1+i\delta},\\
    \Gamma_3&=\gamma_2\frac{2\sqrt{-2\eta_2}}{1+i\delta},\\
    \mathcal{N}&=\frac{2f(iR_1-I_1)}{1+i\delta}.
\end{align}
Equation \eqref{eq:QGTI_2} of the $\mathcal{D}$ field that has been rewritten to
\begin{align}
   \epsilon  \dot{\mathcal{D}}(t)=\gamma\left[-\mathcal{D}(t)+(I_0+\epsilon I_1 \mathcal{D}(t))|\mathcal{E}(t)|^2\right],
\end{align}
has to be discussed as well. Inserting the time derivative operator as well as the nonlinearity leads to
\begin{align}
    \left(\epsilon\partial_0+\epsilon^2\omega_1\partial_0+\epsilon^3(\omega_2\partial_0+\partial_2)\right) \dot{D}(t)
     &=\gamma\left[-D(t)+(I_0+\epsilon I_1 D(t))|E(t)|^2\right].
\end{align}
    \section{Assembling the Full Partial Differential Equation System}
Finally the PDE shall be deduced from the previous results. Therefor, an effective time scale
\begin{align}
    \partial_\theta=\epsilon^2\partial_2+\epsilon^3\partial_3 
\end{align}
is introduced. The results are now included by using $\mathcal{E}=\mathcal{E}_0+\epsilon \mathcal{E}_1$, where higher orders of the field fall into $\mathcal{O}(\epsilon^4)$ due to the definition of the time scale. This leads to:
\begin{align}
    \partial_\theta \mathcal{E}
    =&\epsilon^2\left(\left[-i\delta\frac{\omega_1^2}{2}\partial_0^2-L_2\right]\mathcal{E}+(\gamma_1+\epsilon\gamma_2)\frac{2\sqrt{-2\eta_2}}{1+i\delta}\right)\nonumber\\
    &+\epsilon^3\left(\frac{1-3\delta^2}{12}\omega_1^3\partial_0^3\mathcal{E}+[\mathcal{N}\mathcal{D}-\omega_1L_2]\mathcal{E}+\omega_1\gamma_1\frac{2\sqrt{-2\eta_2}}{1+i\delta}\right).
\end{align}
Next, space can be rescaled as $\epsilon|\omega_1|\partial_0\rightarrow\partial_\sigma$, where the absolute value leads to a switch in sign of the third order dispersion:
\begin{align}
    \partial_\theta \mathcal{E}=&\left[\frac{-i\delta}{2}\partial_\sigma^2-\epsilon^2(-\eta_2+2W_1-i\theta_2)\right]\mathcal{E}+h\mathcal{Y}_0\frac{\sqrt{1-\eta^2}}{1+i\delta}-\frac{1-3\delta^2}{12}\partial_\sigma^3\mathcal{E}\nonumber\\
    &+\epsilon^3\left[\frac{2f(iR_1-I_1)}{1+i\delta}\mathcal{D}-\omega_1(-\eta_2+2W_1-i\theta_2)\right]\mathcal{E}+\epsilon\omega_1h\mathcal{Y}_0\frac{\sqrt{1-\eta^2}}{1+i\delta}+\mathcal{O}(\epsilon^4).
\end{align}
Applying the parameter expansions yields $\epsilon^2(\eta_2+i\theta)+\mathcal{O}(\epsilon^4)=\eta e^{i\theta}-1$. Also, $\epsilon^2f \rightarrow f$ is reset and small terms containing $\epsilon\omega_1$ are removed, yielding
\begin{align}
    \partial_\theta \mathcal{E}=&\left[\frac{-i\delta}{2}\partial_\sigma^2-2f\frac{I_0-iR_0}{(1+i\delta)^2}-1+\eta e^{i\theta}\right]\mathcal{E}+h\mathcal{Y}_0\frac{\sqrt{1-\eta^2}}{1+i\delta}\nonumber\\
    &-\frac{1-3\delta^2}{12}\partial_\sigma^3\mathcal{E}+\epsilon\frac{2f(iR_1-I_1)}{1+i\delta}\mathcal{D}\mathcal{E}+\mathcal{O}(\epsilon^4).
\end{align}
One can progress by multiplying both sides of the equation by $\epsilon^\frac{1}{2}$ and inverting the field's scaling with $\mathcal{E}\epsilon^\frac{1}{2}\rightarrow E$, $\mathcal{Y}_0\epsilon^\frac{1}{2}\rightarrow Y_0$ and $\mathcal{D}\epsilon\rightarrow D$, which results in
\begin{align}
    \partial_\theta E=&\frac{-i\delta}{2}\partial_\sigma^2E-\frac{1-3\delta^2}{12}\partial_\sigma^3E+\frac{2f(iR_1-I_1)}{1+i\delta}DE\nonumber\\&
    +\left[-2 f\frac{I_0-iR_0}{(1+i\delta)^2}-1+\eta e^{-i2\arctan(\delta)}\right]E
    + hY_0\frac{\sqrt{1-\eta^2}}{1+i\delta}.
\end{align}
Here, the original fields from Eqs. \eqref{eq:QGTI_1} - \eqref{eq:QGTI_chi} are recovered. This is necessary to compare the derived PDE with the underlying DAE system.\\
After expanding the exponential function for small values of $\delta$ only the coefficients $R_0$, $I_0$, $R_1$, and $I_1$ remain. Hence, the approximation for the susceptibility shall be finally inserted leading to
\begin{align}
    \partial_\theta E=&\frac{-i\delta}{2}\partial_\sigma^2E-\frac{1-3\delta^2}{12}\partial_\sigma^3E+\frac{4f}{\pi}\frac{(1+iu)(1-i\delta)}{(1+u^2)(1+\delta^2)}DE\nonumber\\
    &+\left[-\frac{2f}{\pi}(1-i\delta)^2\frac{i\ln\sqrt{u^2+1}+\frac{\pi}{2}+\arctan(u)}{\left(1+\delta^2\right)^2}-1+\eta-i2\eta\arctan(\delta)\right]E\nonumber\\
    &+ hY_0\frac{\sqrt{1-\eta^2}}{1+i\delta},
\end{align}
which is this section's final PDE for $E$. The second equation
\begin{align}
    (\epsilon\partial_0+\epsilon^2\omega_1\partial_0+\epsilon^3(\omega_2\partial_0+\partial_2)) \mathcal{D}(t)
     &=\gamma\left[-\mathcal{D}(t)+(I_0+\epsilon I_1 \mathcal{D}(t))|\mathcal{E}(t)|^2\right]
\end{align}
will be simplified using $\epsilon|\omega_1|\partial_0=\partial_\sigma$ as well as $\epsilon^2\partial_2+\omega^3\partial_3=\partial_\theta$. However, one needs to keep in mind that the fields involved still have to be rescaled. Thus, $\mathcal{E}\epsilon^\frac{1}{2}\rightarrow E$ and $\mathcal{D}\epsilon\rightarrow D$ are applied. This leads to
\begin{align}
    \left(\frac{1}{-\omega_1}\partial_\sigma-\epsilon(\partial_\sigma+\partial_\theta)+\epsilon^3\omega_2\partial_0+\omega^4\omega_3\partial_0\right) \frac{1}{\epsilon} D
     &=\gamma\left[\frac{-1}{\epsilon} D+(I_0+ I_1\frac{\epsilon}{\epsilon} D)\frac{1}{\epsilon}|E|^2\right],
 \end{align}
 which is equivalent to
 \begin{align}\left(\frac{1}{-\omega_1}\partial_\sigma-\epsilon\partial_\sigma+\epsilon\partial_\theta-\epsilon^2\omega_1\partial_\sigma-\epsilon^3\omega_1^2\partial_\sigma\right) \frac{1}{\epsilon}D&=
     \frac{\gamma}{\epsilon}\left[-D+(I_0+I_1D)|E|^2\right].
\end{align}
Here, the relations $\omega_2=\omega_1^2$ and $\omega_3=\omega_1^3$ have been used to rewrite all derivative operators with the new scales $\sigma$ and $\theta$.
Now each side is multiplied by $\epsilon$. In a first attempt the lowest order can be used to describe the dynamics of $D$ as
\begin{align}
          \partial_\sigma D
     =-\omega_1\gamma\left[-D+(I_0+I_1D)|E|^2\right].
\end{align}
However, interpreting $\sigma$ as a quasi-spatial variable and $\theta$ as the time variable this is not a differential equation in time $\sigma$. Indeed this section leads to a PDE in $E(t)$ with a constraint for the field $D(t)$. The resulting system is

\begin{align}
    \partial_\theta E=&\frac{-i\delta}{2}\partial_\sigma^2E-\frac{1-3\delta^2}{12}\partial_\sigma^3E+\frac{4f}{\pi}\frac{(1+iu)(1-i\delta)}{(1+u^2)(1+\delta^2)}DE\nonumber\\
    &+\left[-\frac{2f}{\pi}(1-i\delta)^2\frac{i\ln\sqrt{u^2+1}+\frac{\pi}{2}+\arctan(u)}{\left(1+\delta^2\right)^2}-1+\eta-i2\eta\arctan(\delta)\right]E\nonumber\\
    &+ hY_0\frac{\sqrt{1-\eta^2}}{1+i\delta},\\
  0=&-\partial_\sigma D
     +\frac{\gamma}{1+\delta^2}\left[-2D+\left(1+\frac{2}{\pi}\arctan(u)-\frac{4}{\pi}\frac{1}{u^2+1}D\right)|E|^2\right].
\end{align}
The equations are finally simplified to
\begin{align}
    \partial_\theta E &= \left(-i \tilde{d}_2\partial_\sigma^2+\tilde{d}_3\partial_\sigma^3+\mathcal{L}\right)E+\mathcal{N}DE+hY_0\frac{\sqrt{1-\eta^2}}{1+i\delta},\label{eq:PDE1}\\
    0&=\left(\zeta_3\partial_\sigma-2-\zeta_2|E|^2\right)D+\zeta_1|E|^2,\label{eq:PDE2}
\end{align}
with 
\begin{align}
    \tilde{d}_2&=\frac{\delta}{2},\\
    \tilde{d}_3&=\frac{3\delta^2-1}{12},\\
    \mathcal{L}&=-\frac{2f}{\pi}(1-i\delta)^2\frac{i\ln\sqrt{u^2+1}+\frac{\pi}{2}+\arctan(u)}{\left(1+\delta^2\right)^2}-1+\eta-i2\eta\arctan(\delta),  \\
    \mathcal{N}&=\frac{4f}{\pi}\frac{(1+iu)(1-i\delta)}{(1+u^2)(1+\delta^2)},\\
    \zeta_1&=\left(1+\frac{2}{\pi}\arctan(u)\right),\\
    \zeta_2&=\frac{4}{\pi}\frac{1}{u^2+1},\\
    \zeta_3&=-\frac{1+\delta^2}{\gamma}.
\end{align}

In the letter we used an scaled $\sigma$ scale to allow for a better comparison with previous results. Using $\sigma\rightarrow-\omega_1\sigma=\frac{2}{1+\delta^2}\sigma$ leads to
\begin{equation}
 \partial_\theta E = \left(-i\,d_2\,\partial_\sigma^2+d_3\,\partial_\sigma^3\right)E+\left(\mathcal{L}+\mathcal{N}D\right)\,E+h\,Y_0\frac{\sqrt{1-\eta^2}}{1+i\delta}\,.\label{eq:PDE1}
\end{equation}
Here, the values of the coefficients $d_2$ and $d_3$ of the linear spatial
operator in Eq.~\eqref{eq:PDE1} read
\begin{equation}
 d_2=\frac{2\,\delta}{(1+\delta^2)^2}\,,\qquad d_3=\frac{2}{3}\frac{3\,\delta^2-1}{(1+\delta^2)^3}\,,
\end{equation}
Additionally rewriting the equations as a function of $\chi(0)$ and $\chi^\prime(0)$ leads to the coefficients $\mathcal{L}$ and $\mathcal{N}$
\begin{align}
    \mathcal{L}=&\,\eta-1-2\,\eta\,\arctan(\delta)\,i+2\,f\,\left(\frac{1-i\delta}{1+\delta^2}\right)^2\chi_0\,i,  \nonumber\\
    \mathcal{N}=&\,2\,f\,\frac{1-i\delta}{1+\delta^2}\,\chi^\prime_0\,i\,, \label{eq:PDE_coef}
\end{align}
In that case, the equation for the carrier density $D$ takes the form
\begin{equation}
 \gamma^{-1}\,\partial_\sigma D=-D+\Im{\left(\chi_0+\chi^\prime_0D\right)}|E|^2\,.\label{eq:PDE2},
\end{equation}
which is the form presented in the letter.